\documentclass[10.5pt]{IEEEtran}
\usepackage{graphicx}
\usepackage{amsmath}
\usepackage{multicol}
\usepackage{multirow}
\usepackage{stfloats}
\usepackage{booktabs}
\usepackage{mathrsfs}

\usepackage{enumerate}
\usepackage{amssymb}
\usepackage{algorithm}
\usepackage{algpseudocode}

\usepackage{array}
\usepackage{underscore}
\usepackage{breqn}
\usepackage{url}

\usepackage{hyperref}
\hypersetup{hidelinks}
\usepackage{flushend}

\usepackage[square, comma, sort&compress, numbers]{natbib}
\usepackage{graphicx}
\usepackage{epstopdf}
\usepackage{caption}
\usepackage{diagbox}

\usepackage{caption}
\usepackage{subfigure}
\usepackage{color}

\usepackage[outerbars,color]{changebar}
\ifx\pdfoutput\undefined
\else\ifnum\pdfoutput>0
  \usepackage{pdfcolmk}
\fi\fi
\cbcolor{red}
\usepackage{fancyhdr}

\makeatletter
\newcommand{\Rmnum}[1]{\expandafter\@slowromancap\romannumeral #1@}
\makeatother

\usepackage{array}  \newcommand{\PreserveBackslash}[1]{\let\temp=\\#1\let\\=\temp}  \newcolumntype{C}[1]{>{\PreserveBackslash\centering}p{#1}}  \newcolumntype{R}[1]{>{\PreserveBackslash\raggedleft}p{#1}}  \newcolumntype{L}[1]{>{\PreserveBackslash\raggedright}p{#1}}

\ifCLASSINFOpdf
\else
\fi
\begin{document}

\title{\begin{huge}End-to-End Transmission Analysis of Simultaneous Wireless Information and Power Transfer using Resonant Beam\end{huge}}

\author{Wen Fang,
Hao Deng,
Qingwen Liu,~\IEEEmembership{Senior Member,~IEEE},
Mingqing Liu,\\
Mengyuan Xu,
Liuqing Yang,~\IEEEmembership{Fellow, IEEE},
and Georgios B. Giannakis,~\IEEEmembership{Fellow,~IEEE}

\thanks{Manuscript received October 16, 2020; revised April 22, 2021; accepted June 9, 2021. The work was supported by the National Key R\&D Program of China under Grant 2020YFB2103900 and Grant 2020YFB2103902. It was also supported by the National Natural Science Foundation of China under Grant 61771344 and Grant 62071334. (\textit{Corresponding author: Qingwen Liu.})}
		
\thanks{W.~Fang, Q. Liu, M. Liu, and M. Xu are with the College of Electronic and Information Engineering, Tongji University, Shanghai 200000, China (e-mail: wen.fang@tongji.edu.cn, qliu@tongji.edu.cn, clare@tongji.edu.cn, and xumy@tongji.edu.cn).}

\thanks{H. Deng is with the School of Software Engineering, Tongji University, Shanghai 200000, China (e-mail: denghao1984@tongji.edu.cn).}


\thanks{L. Yang and G. B. Giannakis are with the Department of Electrical and Computer Engineering, University of Minnesota, Minneapolis, MN 55455, USA (e-mail: qingqing@umn.edu and georgios@umn.edu).}

}

\maketitle

\begin{abstract}
\normalsize
Integrating the wireless power transfer (WPT) technology into the wireless communication system has been important for operational cost saving and power-hungry problem solving of electronic devices. In this paper, we propose a resonant beam simultaneous wireless information and power transfer (RB-SWIPT) system, which utilizes a gain medium and two retro-reflecting surfaces to enhance and retro-reflect energy, and allows devices to recharge their batteries and exchange information from the resonant beam wirelessly. To reveal the SWIPT mechanism and evaluate the SWIPT performance, we establish an analytical end-to-end (E$2$E) transmission model based on a modular approach and the electromagnetic field propagation. Then, the intra-cavity power intensity distribution, transmission loss, output power, and E$2$E efficiency can be obtained. The numerical evaluation illustrates that the exemplary RB-SWIPT system can provide about 4.20W electric power and 12.41bps/Hz spectral efficiency, and shorter transmission distance or larger retro-reflecting surface size can lead to higher E$2$E efficiency. The RB-SWIPT presents a new way for high-power, long-range WPT, and high-rate communication.
\end{abstract}


\IEEEpeerreviewmaketitle

\section{Introduction}\label{Section1}
With the rapid development of wireless communication technologies, the problem of energy consumption has become prominent \cite{chen2010improved}. The energy consumption of conventional communication devices relies heavily on rechargeable or replaceable batteries, which are expensive and the mobility of the device is limited. Recently, several renewable energy resources, such as solar, wind, have been integrated into communication networks for power supply reference. However, the intermittent and unpredictable nature of these energy sources limits the application and quality-of-service (QoS) of energy harvesting (EH) technologies \cite{krikidis2014simultaneous}.

Wireless power transfer (WPT) is one of the EH technologies that overcome above limitations, which can provide enduring charging for communication devices. The prospect of integrating WPT with communication networks creates a need for technology that can transfer both information and energy simultaneously to devices. Therefore, the concept of simultaneous wireless information and power transfer (SWIPT) is first introduced in \cite{varshney2008transporting}, and a capacity-energy function and a coding theorem are also defined. Developing green technologies and reducing the power consumption of devices are two of the eight major requirements for $5$th generation mobile networks ($5$G) systems identified by industry and academia \cite{agiwal2016next}. Thus, in the era of $5$G, the SWIPT technology can be of fundamental importance for energy supply and information exchange among numerous devices \cite{perera2018simultaneous, liu2020big}.

However, the existing long-range SWIPT technologies face short-range, low-power, unsafe and other challenges. For example, electromagnetic coupling SWIPT has the bottleneck of short transmission distance \cite{masotti2021rf}, radio-frequency (RF) and light-emitting diode (LED) lights based SWIPT face the challenge of low transmission power \cite{liu2016swipt, ma2019simultaneous}, and laser-based SWIPT is difficult to achieve high power transmission safely \cite{fakidis20180}.

Recently, Resonant Beam Charging (RBC), also known as Distribute Laser Charging (DLC), was proposed for high-power and long-range WPT, which relies on the spatially separated cavity to generate the resonant beam for transmitting energy over the air \cite{liu2016dlc, fang2018}. The structure and characteristics of the RBC system were illustrated in \cite{liu2016dlc}, while the process of power transfer in the RBC system was studied in \cite{fang2018, zhang2018distributed2}. \cite{wang2018channel} experimentally demonstrates that it is possible to realize long-range WPT in the RBC system. Furthermore, the RBC technology is identified as one of the innovations for power supply in $6$th generation mobile networks ($6$G) \cite{david20186g}. Afterwards, \cite{xiong2019resonant} and \cite{chen2019resonant} illustrate that the structure of the RBC system can also be used as a communication system to achieve high-rate, high-capacity communication.

\begin{figure}[!t]
    \centering
    \includegraphics[scale=0.32]{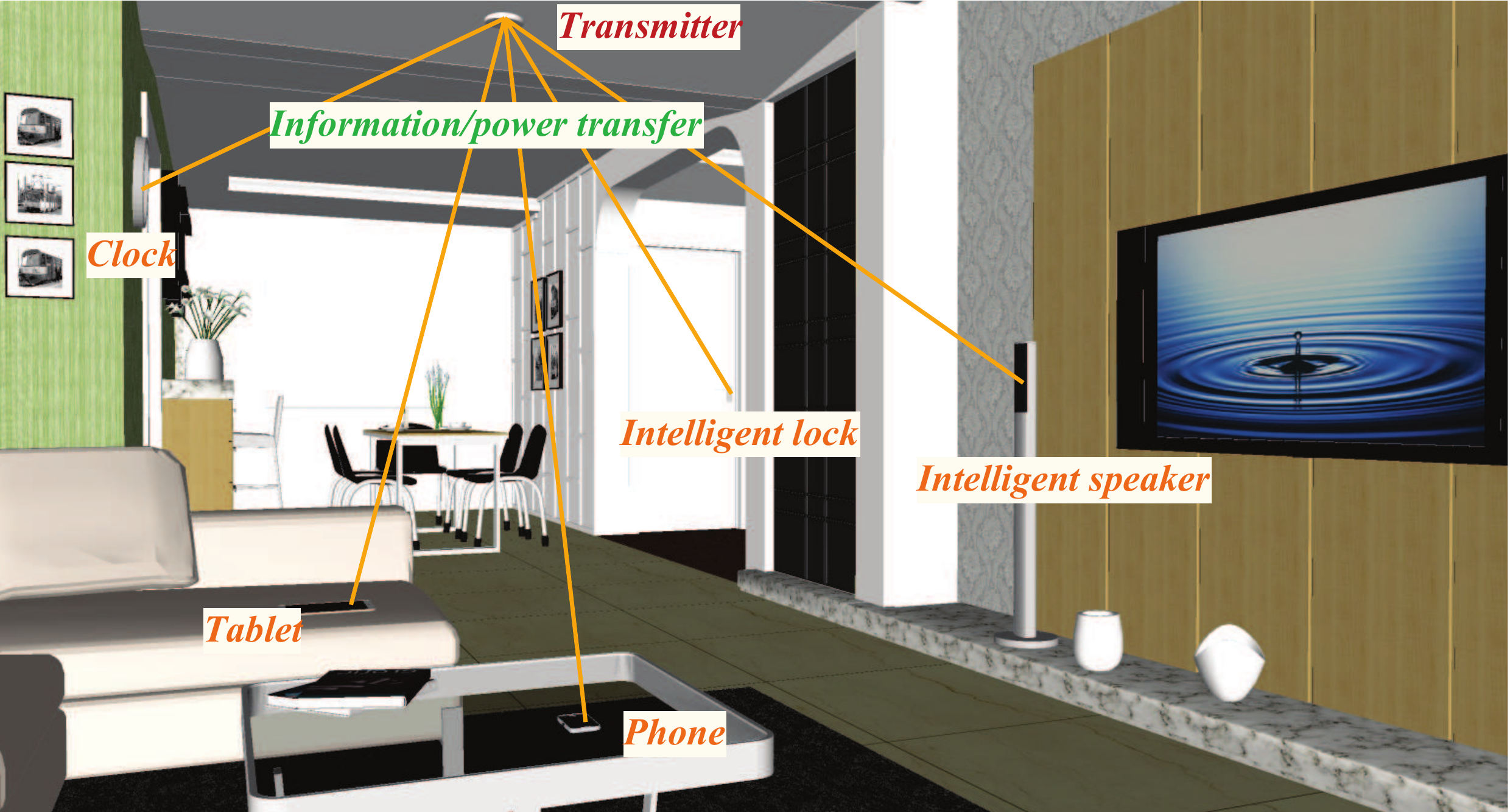}
    \caption{A typical application scenario of the RB-SWIPT system.}
    \label{Fig1}
\end{figure}

Compared with the existing SWIPT technologies, as the energy carrier, the resonant beam can realize long-range, high-power, safe WPT, and high-rate communication. Thus, we propose the resonant beam simultaneous wireless information and power transfer (RB-SWIPT) system based on the structure of RBC system, in which the resonant beam (intra-cavity laser) is the energy-information carrier. The input and output retro-reflectors are placed at the spatially separated transmitter and receiver and used to retro-reflect the resonant beam. Besides, the gain medium at the transmitter can amplify the resonant beam passing through it to provide energy. Figure \ref{Fig1} illustrates a typical application scenario of the RB-SWIPT system. The end-device, e.g. mobile phones, intelligent speakers and intelligent lock systems, embedded with the SWIPT receiver can be charged and exchanged information wirelessly by the transmitter mounted on the ceiling. Thus, high throughput and energy sustainability can be realized for these devices.

In this paper, we first introduce the structure of RB-SWIPT system. Then, we adopt the electromagnetic field propagation and the modular approach to theoretically analyze the transmission model and accurately calculate transmission loss, output electric power, spectral efficiency, and end-to-end (E$2$E) efficiency. Finally, we evaluate energy and information transfer performance with different system parameters. The contributions of this paper include:

\begin{enumerate}
  \item [\bf c1)] We propose the resonant beam simultaneous wireless information and power transfer (RB-SWIPT) system, which can provide long-range, high-power WPT and high-rate communication simultaneously, and thus can support energy sustainability and high throughput for wireless devices.
  \item [\bf c2)] We establish an analytical end-to-end transmission model using modular approach and electromagnetic field propagation for the RB-SWIPT system. It can reveal the intra-cavity power intensity distribution, transmission loss, output power for energy-information transfer, and E$2$E efficiency numerically.
  \item [\bf c3)] The numerical evaluation illustrates that the exemplary RB-SWIPT system can provide about $4.20$W output electric power and $12.41$bps/Hz spectral efficiency.
\end{enumerate}

In the rest of this paper, the RB-SWIPT system structure and E$2$E transmission model will be introduced in Section II. The RB-SWIPT model will be analyzed in Section III modularly. And then, the numerical evaluation of power intensity distribution, transmission loss, output beam power, output electric power, spectral efficiency, and E$2$E efficiency will be described in Section IV. Finally, the paper will be concluded along with the future research in the RB-SWIPT system.

\section{RB-SWIPT System}\label{Section2}
The RB-SWIPT system is able to transfer power and information over a long-range owing to the spatially separated laser cavity structure. In this section, we first introduce the structure of the RB-SWIPT system. Then, the E$2$E SWIPT transmission model is presented summarily.

\subsection{System Overview}
\begin{figure}[!t]
	\centering
    \includegraphics[scale=0.52]{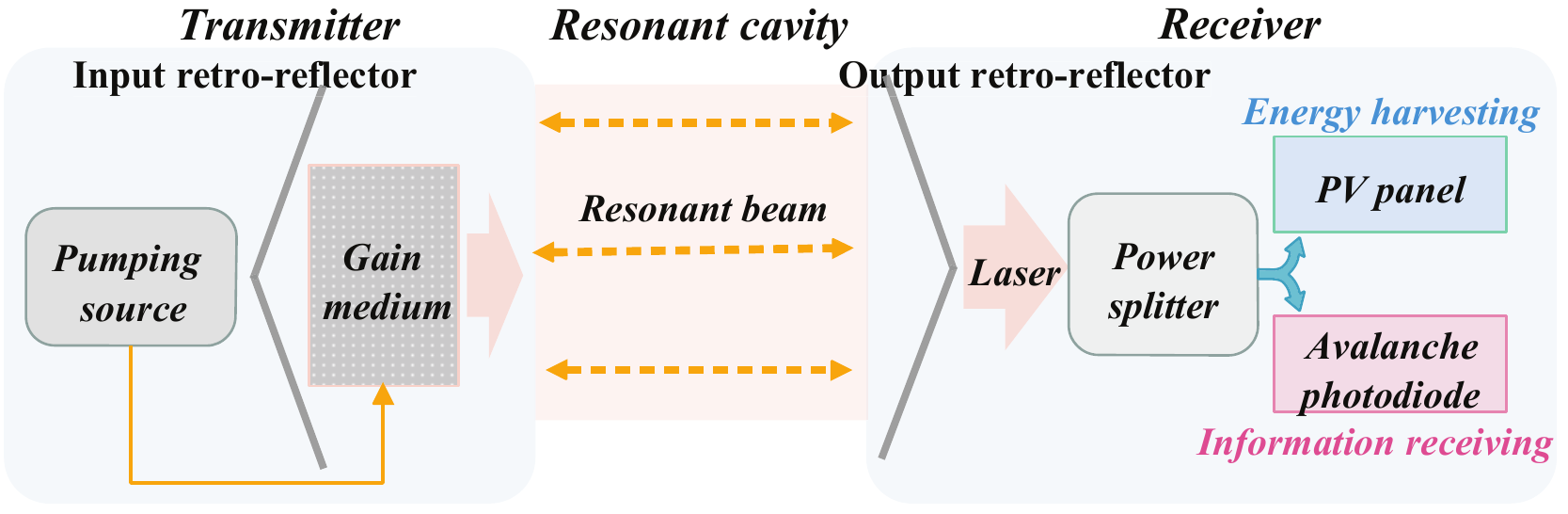}
	\caption{The RB-SWIPT system structure.}
    \label{Fig2}
\end{figure}

The structure of the RB-SWITP system is depicted in Fig. \ref{Fig2}. An RB-SWITP system consists of a transmitter and a receiver, which are spatially separated. The transmitter includes a retro-reflector with $100\%$ reflectivity and a gain medium, and the receiver is comprised of a retro-reflector with partial reflectivity, a power splitter, a photovoltaic (PV) panel and an avalanche photodiode (APD). There exists resonant beam between the transmitter and receiver for energy and information transfer.

\begin{figure}[!t]
	\centering
    \includegraphics[scale=0.53]{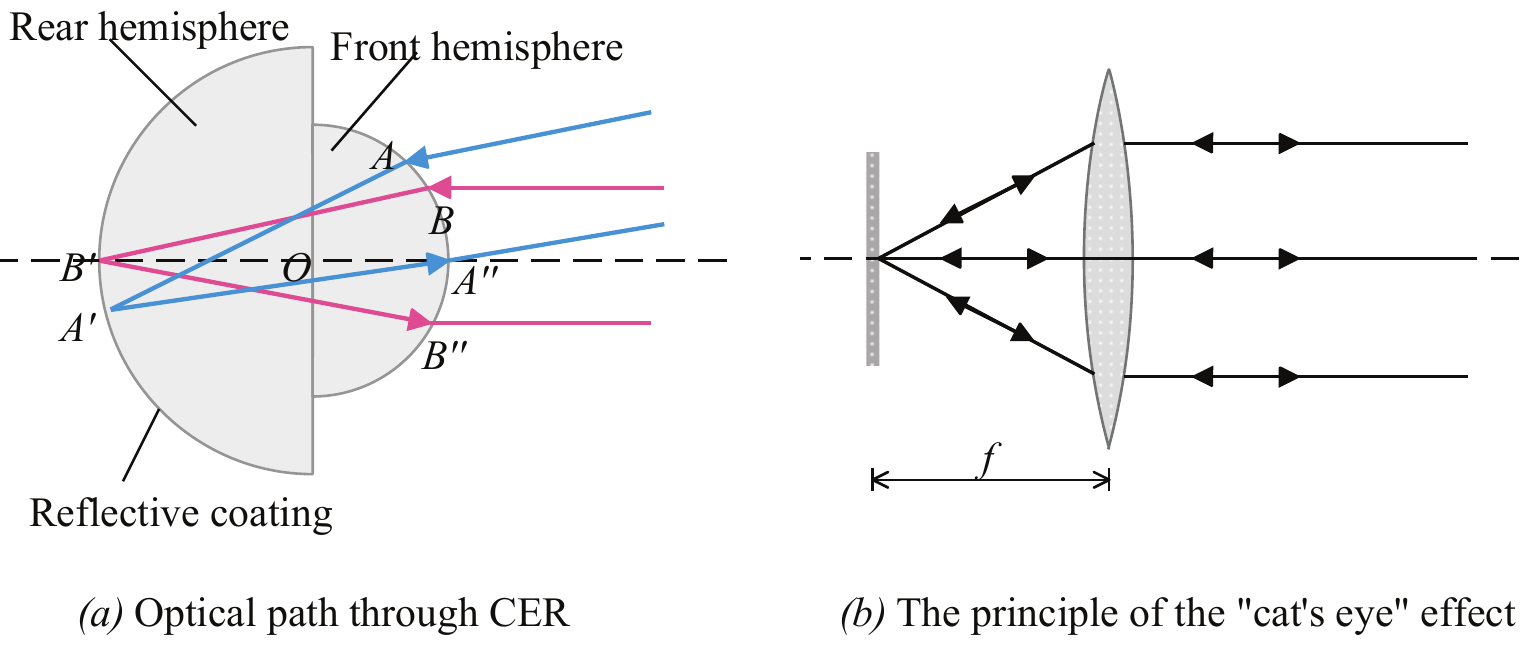}
	\caption{The model of cat's eye retro-reflector.}
    \label{FigCat}
\end{figure}

The energy transfer process in the RB-SWIPT system is as follows. i) In the transmitter, the pumping source provides the pumping electric power to stimulate the gain medium for generating the resonant beam. ii) The gain medium is stimulated to perform atomic transitions, and then form population inversions between the high and low energy levels, thereby pump out the resonant beam \cite{hodgson2005laser}. iii) The pump beam power transmits from the transmitter to the receiver over the air carried by resonant beam. iv) The output retro-reflector receives the beam power. Based on the reflectivity of output retro-reflector, part of beam passes through the retro-reflector forming the output beam, and the rest is reflected to the gain medium for amplification. For example, if the reflectivity of output retro-reflector is 90\%, 10\% of the received energy passes through the output retro-reflector to charge and communicate, and the remaining 90\% is reflected to the transmitter. v) Finally, the power splitter distributes output power to PV panel converting into charging power and APD receiving information.

\begin{figure*}[!t]
	\centering
    \includegraphics[scale=0.65]{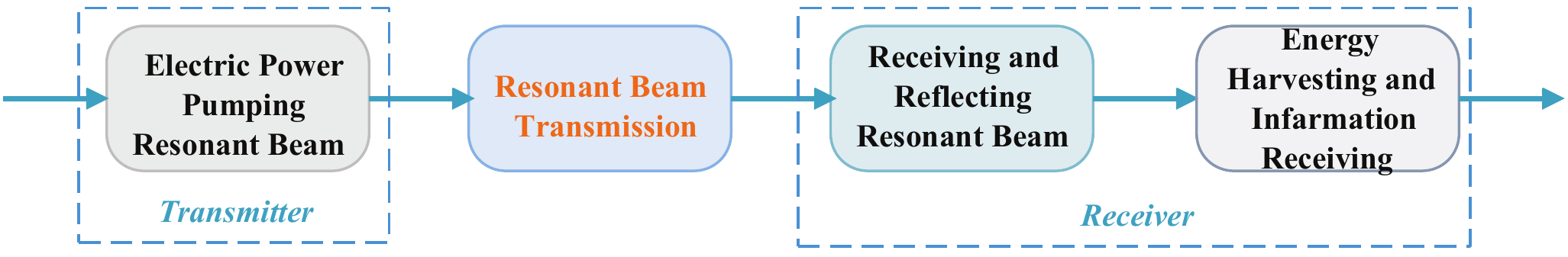}
	\caption{RB-SWIPT end-to-end transmission model.}
    \label{Fig3}
\end{figure*}

For the input and output retro-reflectors, regardless of the direction of the beam incident on the retro-reflector, the beam can be reflected towards the original direction. This property is called retro-reflection. The input/output retro-reflector can be Corner-Cube retro-reflector or cat's eye retro-reflector (CER) \cite{lim2019wireless}.

In this paper, we investigate the RB-SWIPT system with the cat's eye retro-reflector. As depicted in Fig. \ref{FigCat}(a), a CER is generally composed of two hemispheres of different radii, which are glued together centrifugally. Also, the front and rear hemispheres are centered at point $O$. To guarantee the retro-reflective property of CER, it must require the focal point of the image on the front hemisphere to fall on the surface of the rear hemisphere, and the surface of the rear hemisphere coated with a reflective film. For any near-axis beam, it will fall on the sphere of the rear hemisphere after the focus of the front hemisphere. Then, it will be reflected from the sphere of the rear hemisphere, which is equivalent to that a beam is emitted from the focal point of the rear hemisphere. The emitted beam is parallel to the incident beam, i.e. the CER is retro-reflective \cite{rabinovich2004performance}.

In Fig. \ref{FigCat}(a), the two rays are incident from points $A$ and $B$, reflected by points $A'$ and $B'$, and then refracted from points $A''$ and $B''$ respectively. This characteristic is also known as the ``cat's-eye effect". As depicted in Fig. \ref{FigCat}(b), the cat's eye effect can be simplified to a combined model of a lens and reflecting surface, where $f$ is the focal length of the lens. If the focal length of the lens is small compared to the transmission distance, then the effect of focal length on beam transmission can be ignored, thereby the CER can be simplified as a retro-reflecting surface with equal reflecting aperture \cite{snyder1975paraxial}. Thus, the resonant cavity of RB-SWIPT system can be equivalent to a Fabry-P\'{e}rot (FP) cavity \cite{poirson1997analytical}.

\subsection{End-to-End Transmission Model}
As shown in Fig. \ref{Fig3}, the E$2$E energy-information transfer process in the RB-SWIPT system can be divided into four steps: i) electric power pumping resonant beam, ii) resonant beam transmission, iii) receiving and reflecting resonant beam, and iv) energy harvesting and information receiving.

Among them, the electric power provides the energy for gain medium to stimulating out the resonant beam, so the efficiency of the ``electric power pumping resonant beam" can be decided by the physical characteristic of the gain medium and the cross-sectional area of the pumping beam \cite{wang2018channel}. Then, the resonant beam transmits between the transmitter and the receiver. During the power transmission, due to the limited sizes of the input and output retro-reflectors and the existence of the deflection, diffraction, or other loss factors, there must be loss during the resonant beam transmission. Afterwards, the output retro-reflector receives the resonant beam. A portion of beam energy is reflected to the transmitter and the rest is devoted to energy and information transfer \cite{eftekharnejad2012impact}. Finally, the output beam energy is converted into electric power and data by PV panel and APD respectively.

Based on the energy-information transfer process, we will establish a resonant beam transmission model based on electromagnetic field propagation for calculating the transmission loss accurately, and establish the end-to-end SWIPT model to study the energy and information transmission process modularly.

\section{End-to-End SWIPT Model}\label{Section2}
In the RB-SWIPT system, the energy-information transmission carrier is the resonant beam. To understand the transmission process, we establish the analytical E$2$E SWIPT model and analyze the transmission process modularly.

\subsection{Electric Power Pumping Resonant Beam}
The pump source provides the input electric power $P_{in}$ to excite the beam out of the gain medium. The process of pumping resonant beam can be summarized as: i) input electric power $P_{in}$ is converted into pumping optical power $P_{gb}$; ii) pumping optical power is absorbed by the gain medium and converted into the available pumping power (i.e. the power in the upper laser level) $P_{av}$; and, iii) $P_{av}$ is used to realize population inversion and pump out the resonant beam.

The efficiency of the pumping process is given by the excitation efficiency $\eta_{excit}$ \cite{hodgson2005laser},
\begin{equation}\label{eqpump}
    \eta_{excit}=\frac{P_{av}}{P_{in}}.
\end{equation}
For the homogeneously broadened lasers, the power in the upper laser level $P_{av}$ is determined by its physical properties,
\begin{equation}\label{eqgian}
    P_{av}=g_0\ell A I_s,
\end{equation}
where $g_0\ell$ is the small-signal gain, $A$ is the cross-sectional area of the gain medium, and $I_s$ is the saturation intensity, which is an attribute of the laser material. That is, $I_s$ represents the peak incident light intensity that can be amplified by the gain medium, and it can be calculated by $I_s = h\upsilon/\sigma\tau$ for a four-level laser system, where $h\upsilon$ represents photon energy, $\sigma$ is emission cross section for stimulated emission, and $\tau$ is spontaneous decay time of the upper laser level \cite{hodgson2005laser}.

\subsection{Resonant Beam Transmission}
The resonant beam is a kind of electromagnetic wave. Thus, we analyze the resonant beam transmission between the transmitter and receiver based on the electromagnetic field propagation.

The Huygens-Kirchhoff principle in the optics shows that each position on the surface of the wave source can be regarded as a new wave source. From these positions (new wave sources), the wavelets are emitted. The field distribution at a position in space is the result of coherent superposition of these wavelets at that position \cite{hodgson2005laser, fox1963modes, lissak1990transverse}. The expression of this theory can be depicted as Fresnel-Kirchoff's diffraction integral:
\begin{equation}\label{eq5}
    u(x,y)=\frac{ik}{4\pi}\iint\limits_S u(x',y')\frac{e^{-ik\rho}}{\rho}(1+cos\theta)ds',
\end{equation}
where $u(x',y')$ is the field distribution function of the wave source point $(x',y')$ on the primary surface, and $u(x,y)$ is the field distribution function of the observation point $(x,y)$. $\rho$ is the distance from $(x',y')$ to $(x,y)$, and $\theta$ represents the angle between the direction determined by two the points $($$(x,y)$ and $(x',y')$$)$ and the unit normal to the wave source surface. In addition, $i$ is the imaginary unit. $k$ is the wave vector and can be calculated by
\begin{equation}\label{eq6}
    k= \frac{2\pi}{\lambda},
\end{equation}
where $\lambda$ represents the beam wavelength. $ds'$ is the surface element at point $(x',y')$ on the wave source surface.

\begin{figure}[!t]
	\centering
    \includegraphics[scale=0.7]{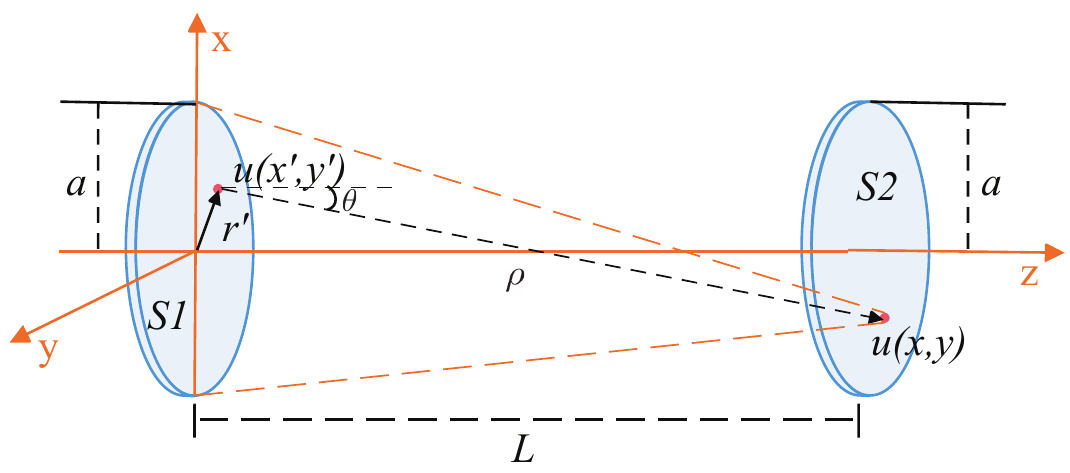}
	\caption{Field analysis in the RB-SWIPT system.}
    \label{Fig4}
\end{figure}

The beam propagation in the resonant cavity is actually the beam spatial diffraction process under the constraints of the retro-reflector sizes. Applying the Fresnel-Kirchoff's diffraction integral to the RB-SWIPT system shown in Fig. \ref{Fig4}: the cavity length is $L$, the radii of the input and output retro-reflecting surfaces are $a$. $r'$ represents the radius inside the surface (i.e. the distance between the center of the retro-reflecting surface $(0,0)$ and any point on the surface $(x,y)$). Besides, as illustrated in Fig. \ref{Fig4}, $r' \leq a$. The optical axes of the surfaces $S1$ and $S2$ coincide. Besides, the direction of optical axis is taken as the z-axis, and the directions perpendicular to the optical axis are used as the x-axis and y-axis to establish a three-dimensional space rectangular coordinate system.

As shown in Fig. \ref{Fig4}, the cavity length $L$ is much greater than the retro-reflector diameter $2a$. The included angle between the propagation direction of paraxial rays and the optical axis is close to zero, so the distance between the positions on the two retro-reflectors can be approximately set as $L$. Therefore, the integral factor $(1+cos\theta)/\rho$ in \eqref{eq5} can be set as $2/L$ approximately, and then it can be removed from the integral sign. And yet, after one propagation, the field distribution function $u_1(x',y')$ on surface $S1$ and the field distribution function $u_2(x,y)$ on surface $S2$ can be simplified as
\begin{equation}\label{eq7}
    u_2(x,y) = \frac{i}{\lambda L}\iint\limits_{S1} u_1(x',y')e^{-ik\rho}ds'.
\end{equation}

\begin{figure}[!t]
	\centering
    \includegraphics[scale=0.65]{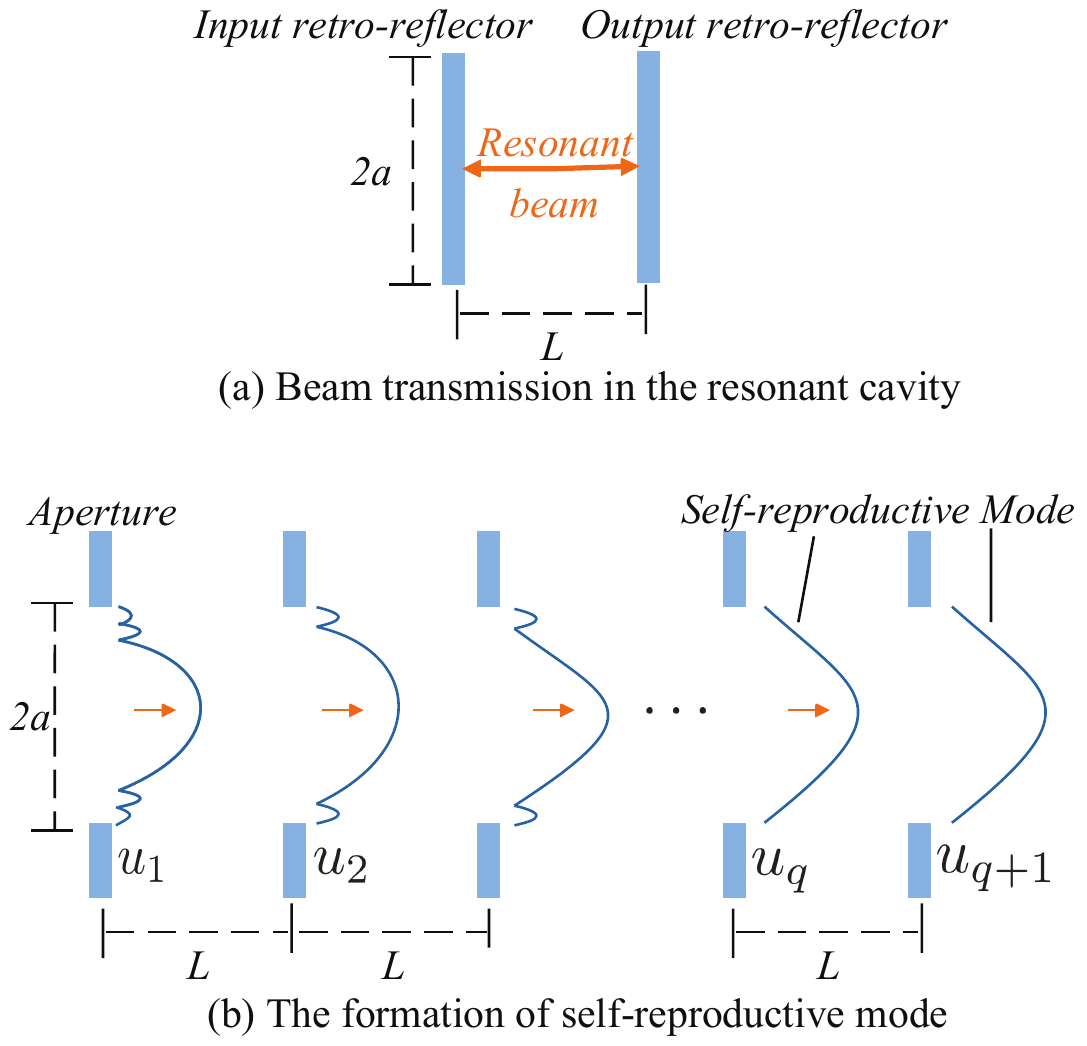}
	\caption{Beam transmission in the resonant cavity (Self-reproductive mode, i.e. Fox-Li numerical iteration method).}
    \label{Fig5}
\end{figure}

Then, after $m$ propagations, the relationship between the field distribution function $u_m(x',y')$ on $S1$ and $u_{m+1}(x,y)$ on $S2$ also satisfies \eqref{eq7}, that is
\begin{equation}\label{eq8}
    u_{m+1}(x,y) = \frac{i}{\lambda L}\iint\limits_{S1} u_m(x',y')e^{-ik\rho}ds'.
\end{equation}

%

In the RB-SWIPT system, the resonant beam is reflected back and forth between two parallel retro-reflecting surfaces, as shown in Fig. \ref{Fig5}(a). This is equivalent to a transmission line comprising a series of apertures as depicted in Fig. \ref{Fig5}(b). Due to the limited geometry sizes of the retro-reflecting surfaces and the existence of loss factors such as absorption, scattering, and deflection, the wavefront of the resonant beam will distort after each aperture propagation, and part of the beam will deviate from the original direction. These deviated beams will be blocked the next time they pass through the aperture. In this way, the above transmission process is repeated in succession through the aperture.

Each time the resonant beam passes the aperture, its amplitude and phase change, and its energy is concentrated towards the optical axis. After multiple round trips, the amplitude distribution of the resonant beam remains essentially constant. Thus, a stable field distribution, which is called a self-reproductive mode, is formed. In the RB-SWIPT system, when the number of beam transmission $m$ is sufficiently large, the field distribution on the retro-reflecting surface S1 is reproduced that on the retro-reflecting surface S2:
\begin{equation}\label{eq9}
    u_{m+1}=\frac{1}{\tau}u_m,
\end{equation}
that is, the field distribution $u_{m+1}$ on the retro-reflecting surface S1 is identical to $u_m$ on the retro-reflecting surface S2 except for a coordinate-independent complex constant factor $\tau$ representing the amplitude attenuation. This field whose amplitude distribution will no longer change is the resonant cavity mode, which can be expressed by the same function $u$,
\begin{equation}\label{eq10}
    u(x,y) = \tau \frac{i}{\lambda L}\iint\limits_{S'} u(x',y')e^{-ik\rho}ds'.
\end{equation}

Introducing an integral kernel associated with two retro-reflectors coordinate systems,
\begin{equation}\label{eq11}
    \Phi(x,y,x',y')=\frac{i}{\lambda L}e^{-ik\rho (x,y,x',y')}.
\end{equation}
Then, \eqref{eq10} can be transformed into an integral eigenfunction with a complex constant eigenvalue,
\begin{equation}\label{eq12}
    u(x,y) = \tau \iint\limits_{S'} \Phi(x,y,x',y') u(x',y')ds'.
\end{equation}


A numerical iterative algorithm for calculating the integral eigenfunction of the resonant cavity is proposed by A. G. Fox and T. Li in 1960, which is called ``Fox-Li" numerical iteration method \cite{fox1961resonant}. The realization of the ``Fox-Li" numerical iteration method relies on the fact that: the beam transmission in the resonant cavity can be simulated by the integral eigenfunction of the resonant cavity. As depicted in Fig. \ref{Fig5}(b), during the resonant beam transmits between the input and output retro-reflecting surfaces, each time the integral is calculated, it's equivalent to one beam transmission process in the cavity. The result of the last calculation is the input of the next calculation. The iteration continues until the diffraction screens out a self-reproductive mode, which is an optical field mode that can exist stably in a resonant cavity. That is, in the presence of a self-reproductive mode in the resonant cavity, the amplitude distribution of the optical field is fully reproducible if the beam diffracts and propagates in one round trip \cite{fox1963modes, fox1961resonant}. As shown in Fig. \ref{Fig5}(b), $u_{q+1}$ can exactly reproduce the amplitude distribution of $u_q$ when the resonant beam propagates enough times.

Afterwards, the beam energy in a plane is the integral of the beam power intensity concerning the plane area. The transmission loss is the difference between the energy on the two planes. Thus, after the resonant beam transmits from the transmitter to the receiver, the power transmission loss $\delta$ can be figured out as \cite{gordon1964equivalence, li1965diffraction}:
\begin{equation}\label{eq16}
    \delta = \frac{\iint\limits_{S}\left|u_1\right| ^2 ds - \iint\limits_{S}\left|u_2\right| ^2 ds}{\iint\limits_{S}\left|u_1\right| ^2 ds}.
\end{equation}

\subsection{Resonant Beam Output}
According to the electromagnetic field propagation method, the beam power intensity at any position in the resonant cavity can be obtained and the beam transmission loss can be calculated by \eqref{eq12} and \eqref{eq16} accurately. The transmission coefficient (i.e. transmission efficiency) $\epsilon$ of the resonant beam can be depicted as:
\begin{equation}\label{eq17}
    \epsilon = 1 - \delta.
\end{equation}

We set the transmission coefficient from the transmitter to the receiver as $\epsilon_1$, and the coefficient from the receiver to the transmitter as $\epsilon_2$. Based on the above analysis, the formula of the output beam power $P_{out}$ can be written as \cite{hodgson2005laser}:
\begin{equation}\label{eq18}
\begin{aligned}
    P_{out} &= A_b I_s \frac{(1-R)\epsilon_1}{1-R\epsilon_1\epsilon_2 + \sqrt{R\epsilon_1\epsilon_2}(\frac{1}{\epsilon_1\epsilon_2 V_s} - V_s)} \\
    & \left[g_0\ell-\left|\ln \sqrt{RV_s^2\epsilon_1 \epsilon_2}\right|\right],
\end{aligned}
\end{equation}
where $A_b$ is the cross-sectional area of the beam, $I_s$ is the saturation intensity of the gain medium, $R$ is the reflectivity of the output retro-reflector, $V_s$ is the transmission coefficient in the gain medium. According to \eqref{eqpump} and \eqref{eqgian}, if the excitation efficiency $\eta_{excit}$ of the gain medium, whose value is decided by the medium's physical characteristics, is known, the small-signal gain can be calculated using:
\begin{equation}\label{eq19}
    g_0\ell=\frac{\eta_{excit}P_{in}}{AI_s},
\end{equation}
where $A$ is the cross-sectional area of the gain medium. Thus, \eqref{eq19} is substituted into \eqref{eq18} to obtain:
\begin{equation}\label{eq20}
\begin{aligned}
    P_{out} &= A_b I_s \frac{(1-R)\epsilon_1}{1-R\epsilon_1\epsilon_2 + \sqrt{R\epsilon_1\epsilon_2}(\frac{1}{\epsilon_1\epsilon_2 V_s} - V_s)}\\
    &\left[\frac{\eta_{excit}P_{in}}{AI_s}-\left|\ln \sqrt{RV_s^2\epsilon_1 \epsilon_2}\right|\right].
\end{aligned}
\end{equation}

\subsection{Energy Harvesting and Information Receiving}
The output beam power is utilized to transfer energy and information. The power splitter in the receiver divides the received beam power into two power streams with a certain power-splitting ratio \cite{perera2018simultaneous}. Let $\gamma$ represent the power-splitting ratio. In addition, the energy stream is converted into electric power through the PV panel, and the information is received by the APD \cite{sera2007pv, cvijetic2008performance}.

\begin{figure}[!t]
    \centering
     \includegraphics[scale=0.51]{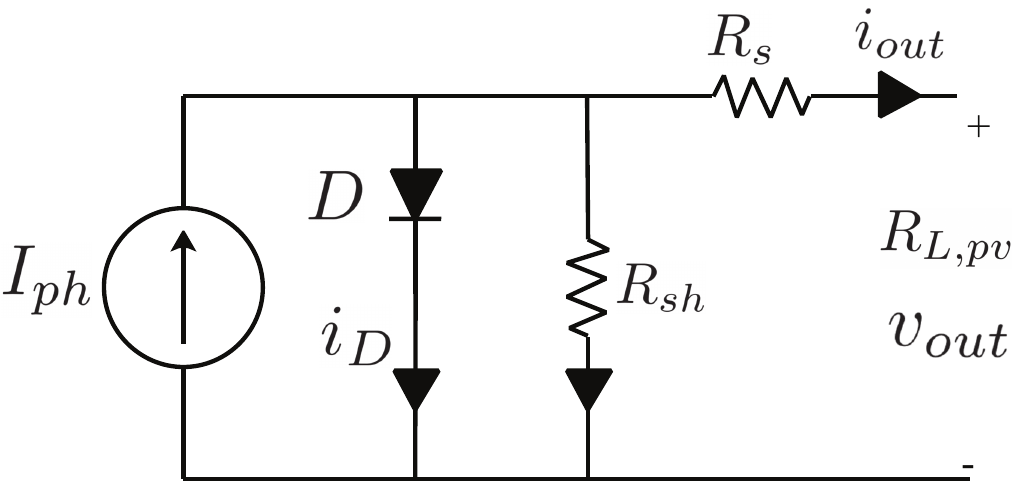}
    \caption{Equivalent circuit of a photovoltaic panel.}
    \label{FigPV}
\end{figure}

\textbf{\textit{D.1) Energy harvesting}}

Figure \ref{FigPV} shows an equivalent circuit of the single-diode model for PV panel. Based on the general current-voltage characteristic of the PV panel, the output current $i_{out}$ is:
\begin{equation}\label{PV}
    i_{out} = I_{ph} - I_o \left[e^{\frac{v_{out} + i_{out} R_s}{n_s V_t}} - 1\right] - \frac{v_{out} + i_{out} R_s}{R_{sh}}.
\end{equation}
In the above equation, $V_t$ is the junction thermal voltage:
\begin{equation}\label{voltage}
    V_t = \frac{FKT}{q},
\end{equation}
where:
\begin{itemize}
  \item $I_{ph}$ - the photo-generated current
  \item $I_{o}$ - dark saturation current
  \item $R_{s}$ - panel series resistance
  \item $R_{sh}$ - panel parallel (shunt) resistance
  \item $F$ - diode quality (ideality) factor
\end{itemize}
are the parameters of the model, while $K$ is the Boltzmann's constant, $q$ is the quantity of electric charge, $n_s$ is the number of cells in the panel connected in series, and $T$ is the temperature in Kelvin. $I_{ph}$ depends on the received beam power of PV panel, which can be expressed as
\begin{equation}\label{current}
    I_{ph} = \gamma \psi P_{out},
\end{equation}
where $\psi$ is the conversion responsivity of PV. Finally, the output electric power for energy supply can be written as:
\begin{equation}\label{outelepower}
\begin{aligned}
      P_{e} &= i_{out} v_{out} = i_{out}^2 R_{L,pv},
\end{aligned}
\end{equation}
where $R_{L,pv}$ is the load resistance of the PV panel.

\textbf{\textit{D.2) Information Receiving}}

The maximum information transfer rate, i.e. spectral efficiency, can be written as \cite{lapidoth2009on}
\begin{equation}\label{SE}
    C = \frac{1}{2}\log \left( 1+\frac{S e}{2 \pi N} \right),
\end{equation}
where $\frac{S}{N}$ represents the signal-to-noise ratio (SNR). $S$ is signal power and $N$ is noise power. The signal power is \cite{lapidoth2009on, perera2018simultaneous}
\begin{equation}\label{SP}
    S = \left[(1-\gamma) P_{out}\eta_{bi}\right]^2,
\end{equation}
where $\eta_{bi}$ is the photo-electric conversion efficiency of APD. $N$ is the power of an additive Gaussian white noise (AGWN), which can be expressed with shot noise $N_{shot}$ and thermal noise $N_{thermal}$ as \cite{lapidoth2009on, perera2018simultaneous}
\begin{equation}\label{NP}
\begin{aligned}
    N & = N_{shot}^2 + N_{thermal}^2 \\
    & = 2q \left\{ \left[(1-\gamma) P_{out}\eta_{bi} \right] + I_{bc} \right\} B_n + \frac{4KTB_n}{R_{L,apd}},
\end{aligned}
\end{equation}
where $q$ is the quantity of electric charge, $I_{bc}$ is the background current, $B_n$ is the noise bandwidth, $K$ is the Boltzmann's constant, $T$ is the temperature in Kelvin, and $R_{L,apd}$ is the load resistor.
\begin{table}[!t]
    \setlength{\abovecaptionskip}{0pt}
    \setlength{\belowcaptionskip}{-3pt}
    \centering
        \caption{Parameters related to the gain medium, the retro-reflecting surface and Fox-Li algorithm \cite{hodgson2005laser}.}
        \vskip .05in
    \begin{tabular}{C{1cm} C{4.3cm} C{1.8cm}}
    \hline
     \textbf{Symbol} & \textbf{Parameter} & \textbf{Value}  \\
    \hline
    \bfseries{$I_s$} & {Saturation intensity} & {$1260W/cm^2$} \\
    \bfseries{$V_s$} & {Transfer coefficient of gain medium} & {$99\%$} \\
    \bfseries{$R$} & {Output retro-reflector reflectivity} & {$95\%$} \\
    \bfseries{$\eta_{excit}$} & {Excitation efficiency} & {$51.48\%$} \\
    \bfseries{$P_{in}$} & {Input electric power} & {$200W$} \\
    \bfseries{$N_{itr}$} & {Fox-Li iteration number} & {$300$} \\
    \hline
    \label{table1}
    \end{tabular}
\end{table}

Furthermore, by varying the power-splitting ratio, the harvested energy and information rate can be balanced based on the system requirements. The end-to-end performance also can be improved by optimizing the transmission loss and the power-splitting ratio.

\begin{table}[!t]
    \setlength{\abovecaptionskip}{0pt}
    \setlength{\belowcaptionskip}{-3pt}
    \centering
        \caption{Parameters of energy harvesting \cite{perales2016characterization}}.
        \vskip .05in
    \begin{tabular}{C{1cm} C{3.5cm} C{2.0cm}}
    \hline
     \textbf{Symbol} & \textbf{Parameter} & \textbf{Value}  \\
    \hline
    \bfseries{$I_o$} & {Dark saturation current} & {$9.89\times10^{-9}A$}\\
    \bfseries{$F$} & {Diode quality factor} & {$1.105$} \\
    \bfseries{$n_s$} & {Number of PV cell} & {$40$} \\
    \bfseries{$R_s$} & {Panel series resistance} & {$0.93\Omega$} \\
    \bfseries{$R_{sh}$} & {Panel parallel resistance} & {$52.6k\Omega$} \\
    \bfseries{$R_{L,pv}$} & {Load resistance} & {$100\Omega$} \\
    \bfseries{$\psi$} & {Conversion responsivity} & {$0.0161A/W$} \\
    \hline
    \label{table2}
    \end{tabular}
\end{table}

\section{Numerical Evaluation}
In this section, to evaluate the SWIPT performance, we numerically analyze the transmission loss, output beam power, output electric power, spectral efficiency, and end-to-end efficiency of the RB-SWIPT system with specified structural parameters.

\subsection{Parameter Setting}
The parameters relevant to the numerical evaluation of SWIPT performance are shown in Table \ref{table1}, \ref{table2} and \ref{table3} \cite{crump2007100, hodgson2005laser}. We adopt the thin crystal composed of Nd:YVO$_4$ as the gain medium, which is attached to the input retro-reflector and has the same radius as the retro-reflector \cite{hodgson2005laser}. The parameters related to the gain medium and the retro-reflector are presented in Table \ref{table1}. For energy harvesting, we adopt a vertical multi-junction PV cell. Table \ref{table2} shows the parameters of the PV panel, which are obtained by using PV equivalent circuit fitting measured data of a real PV cell product \cite{perales2016characterization}. Meanwhile, as shown in Table \ref{table3}, we obtain parameters of APD according to experimental results using APD for information receiving \cite{demir2017handover}, where $q$, $K$ and $T$ are generic. Besides, the parameters related to information receiving $B_n$, $I_{bc}$, and $R_{L,APD}$ are extracted from the corresponding literatures \cite{quintana2017high, moreira1997optical, xu2011impact}. The power-splitting ratio $\gamma$ can be determined based on the charging and communication needs.

\begin{table}[!t]
    \setlength{\abovecaptionskip}{0pt}
    \setlength{\belowcaptionskip}{-3pt}
    \centering
        \caption{Parameters of information receiving.}
        \vskip .05in
    \begin{tabular}{C{1cm} C{3.5cm} C{2.0cm}}
    \hline
     \textbf{Symbol} & \textbf{Parameter} & \textbf{Value}  \\
    \hline
    \bfseries{$\eta_{bi}$} & {Conversion efficiency of APD} & {$0.6A/W$} \cite{demir2017handover} \\
    \bfseries{$q$} & {Quantity of electric charge} & {$1.6\times10^{-19}$} \\
    \bfseries{$B_n$} & {Noise bandwidth} & {$811.7MHz$} \cite{quintana2017high} \\
    \bfseries{$I_{bc}$} & {Background current} & {$5100\mu A$} \cite{moreira1997optical} \\
    \bfseries{$K$} & {Boltzmann constant} & {$1.38\times10^{-23}$} \\
    \bfseries{$T$} & {Kelvin Temperature} & {$300K$} \\
    \bfseries{$R_{L,apd}$} & {Load resistor} & {$10K\Omega$} \cite{xu2011impact} \\
    \hline
    \label{table3}
    \end{tabular}
\end{table}

\begin{figure}[!t]
    \centering
     \includegraphics[scale=0.46]{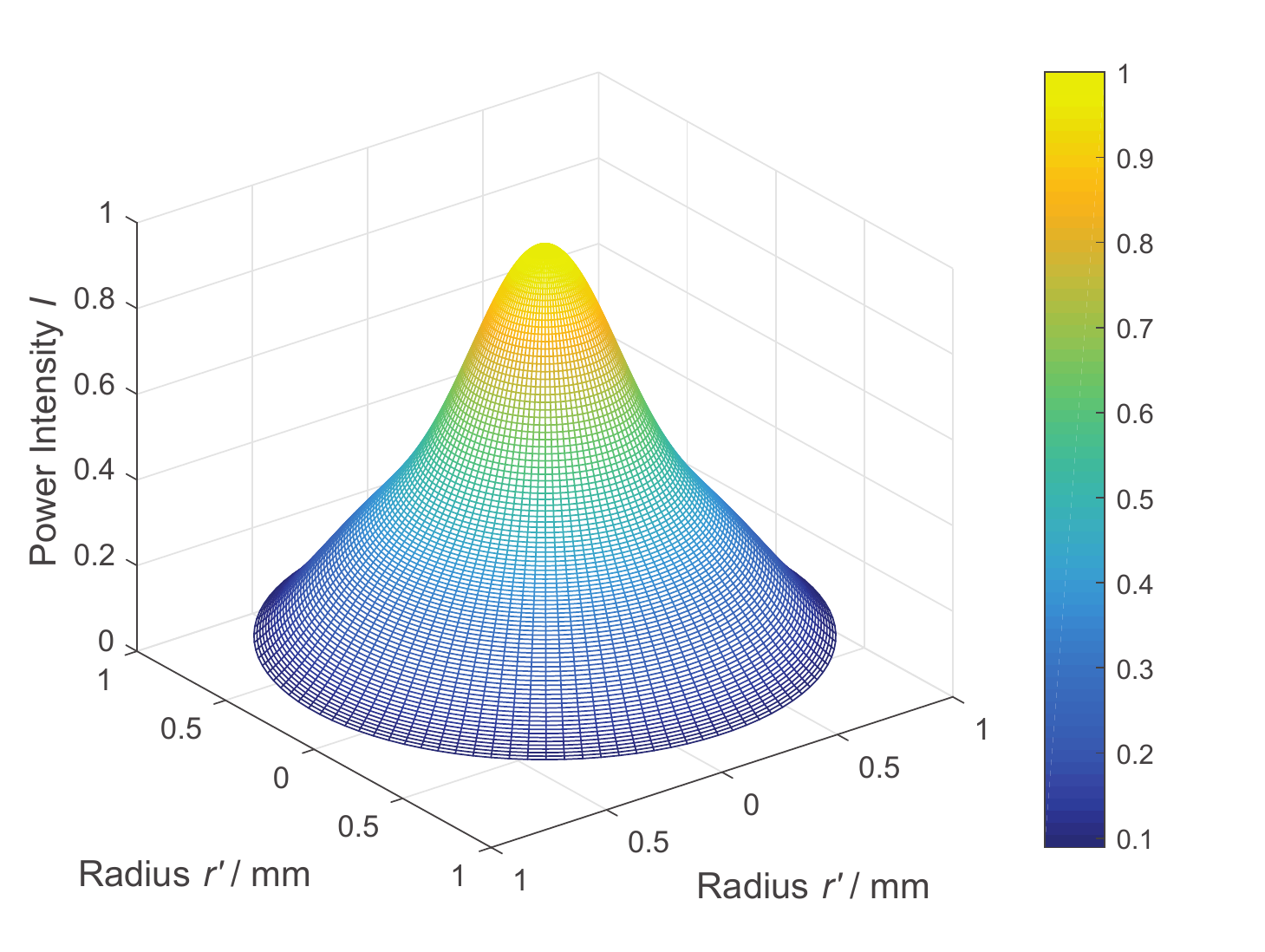}
    \caption{Power intensity distribution on input retro-reflecting surface $S1$}
    \label{Fig6}
\end{figure}

\begin{figure}[!t]
    \centering
     \includegraphics[scale=0.73]{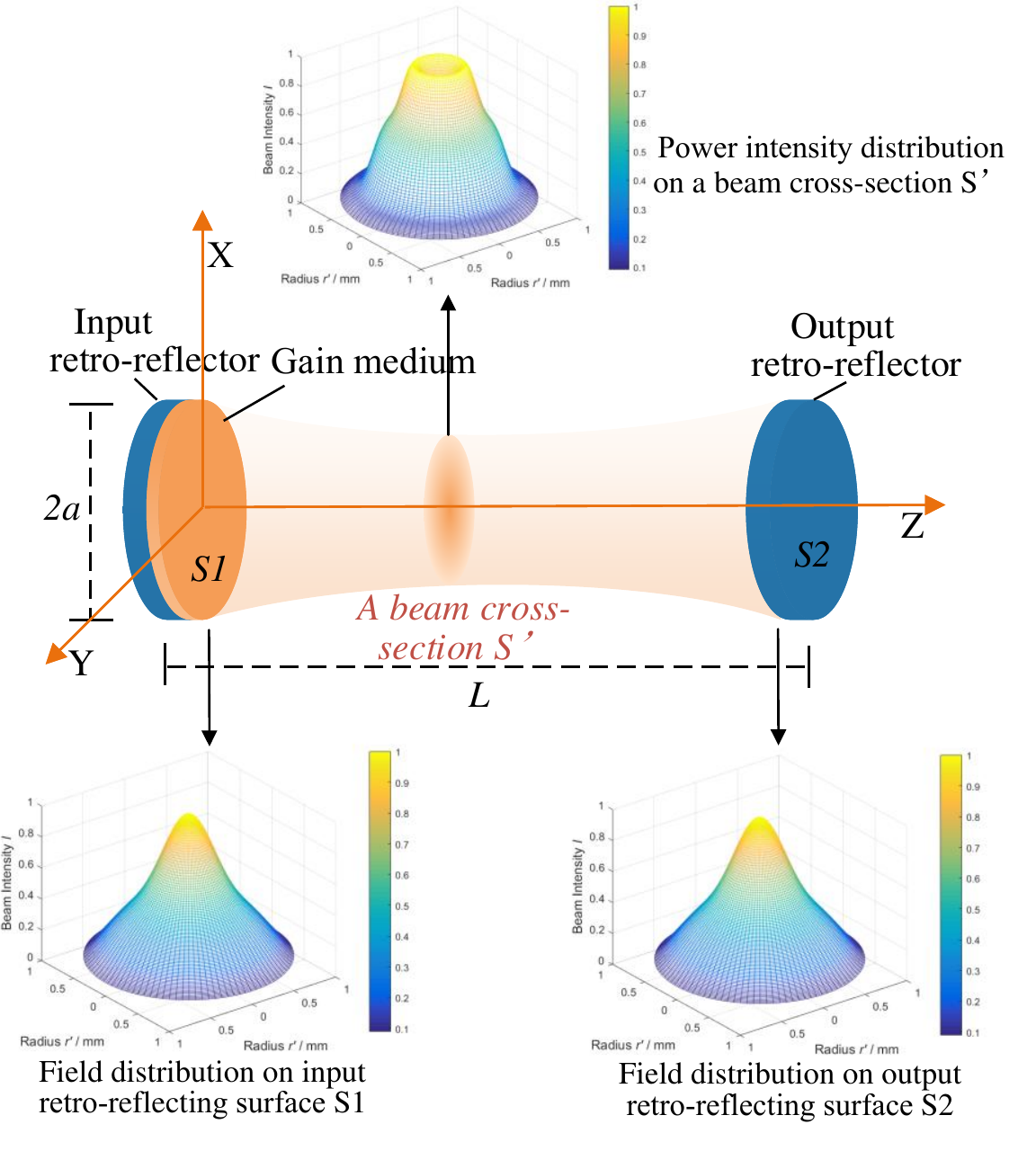}
    \caption{Power intensity at any position in the resonant cavity.}
    \label{Fig8}
\end{figure}

\subsection{Power Intensity Distribution in the Resonant Cavity}
Using the Fresnel-Kirchoff's diffraction integral and ``Fox-Li" numerical iteration method, the power intensity distribution in the RB-SWIPT system shown in Fig. \ref{Fig2} can be calculated. If the cavity length $L$ is $1m$ and the radius of retro-reflector $a$ is $1mm$, the power intensity distribution on the input retro-reflecting surface $S1$ is depicted in Fig. \ref{Fig6}. The power intensity in the center of the surface is the largest, and it gradually decreases radially. Furthermore, the beam power intensity is equivalent at the same radial position on the retro-reflecting surface. When $r'$ is $0.3mm$, the relative power intensity is about 0.7.

Afterwards, based on the calculation of Fresnel-Kirchoff's diffraction integral and ``Fox-Li" numerical iteration, the beam power intensity of any position in the resonant cavity can be obtained, given the cavity length $L$ and the retro-reflector radius $a$. As Fig. \ref{Fig8} shows, if the cavity length $L$ is $1m$ and the radius of retro-reflector $a$ is $1mm$ in an RB-SWIPT system, the resonant beam propagates as a Gaussian beam in the resonant cavity, and the power intensity distribution on any plane in parallel with the retro-reflecting surface (i.e. any beam cross-section) can be obtained. In Fig. \ref{Fig8}, the power intensity distributions on input retro-reflecting surface $S1$, output retro-reflecting surface $S2$, and a beam cross-section $S'$ are depicted. Afterwards, from the power intensity distributions, the power intensity of any point on $S1$, $S2$, and $S'$ can be obtained. Similarly, the power intensity of any point in the resonant cavity can also be obtained through power intensity distribution on the plane parallel to the retro-reflecting surface. On this basis, the transmission loss in the cavity can be calculated in \eqref{eq16}.

\subsection{Transmission Loss and Output Beam Power}

To evaluate the performance of RB-SWIPT system with different system structure parameters, we first analyze the transmission loss in the resonant cavity and the output beam power of the system under different cavity lengths and different radii of retro-reflectors.

\begin{figure}[!t]
    \centering
     \includegraphics[scale=0.58]{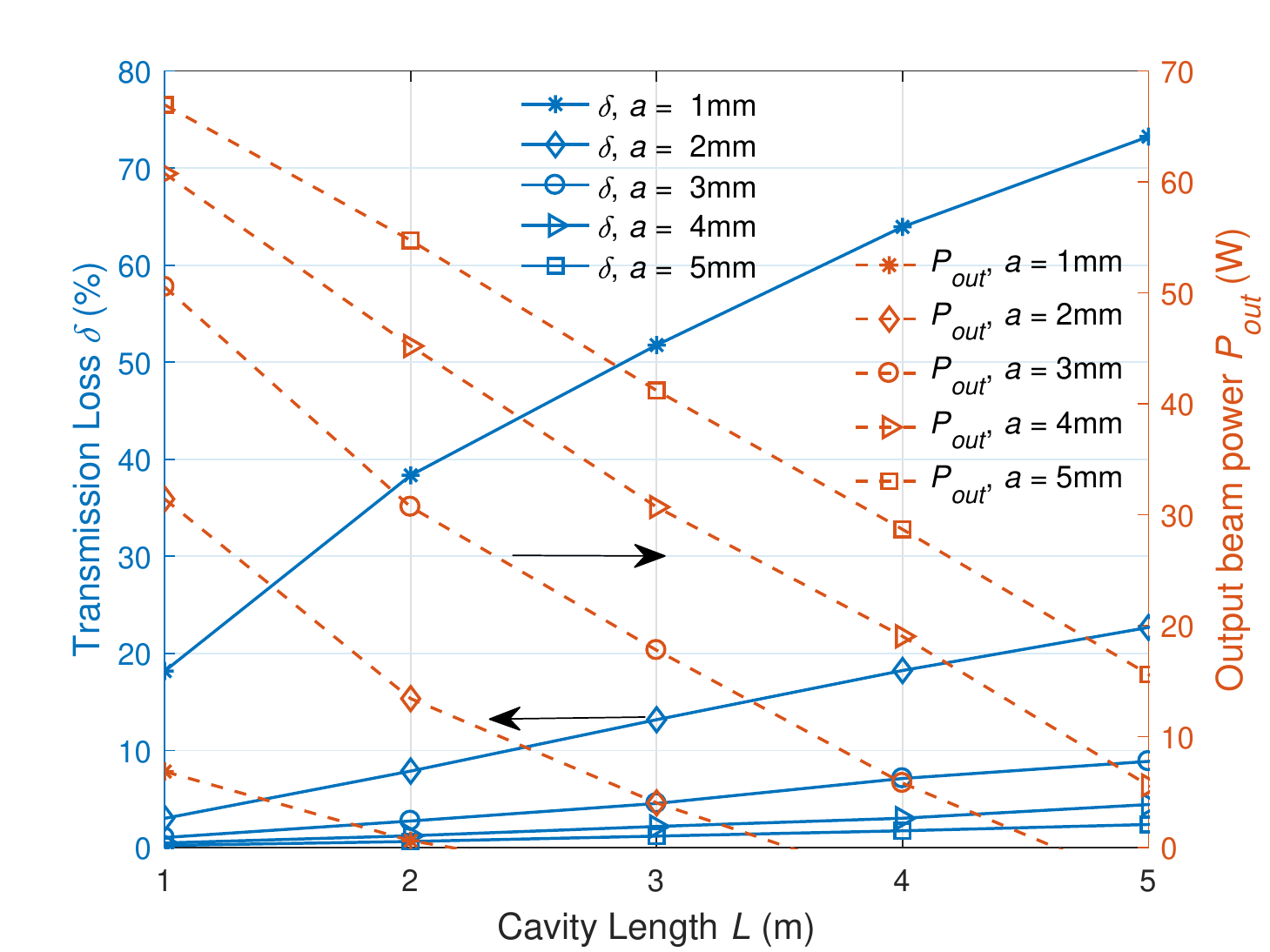}
    \caption{Transmission loss $\delta$ and output beam power $P_{out}$ versus resonant cavity length $L$.}
    \label{Fig9}
\end{figure}

\begin{figure}[!t]
    \centering
     \includegraphics[scale=0.58]{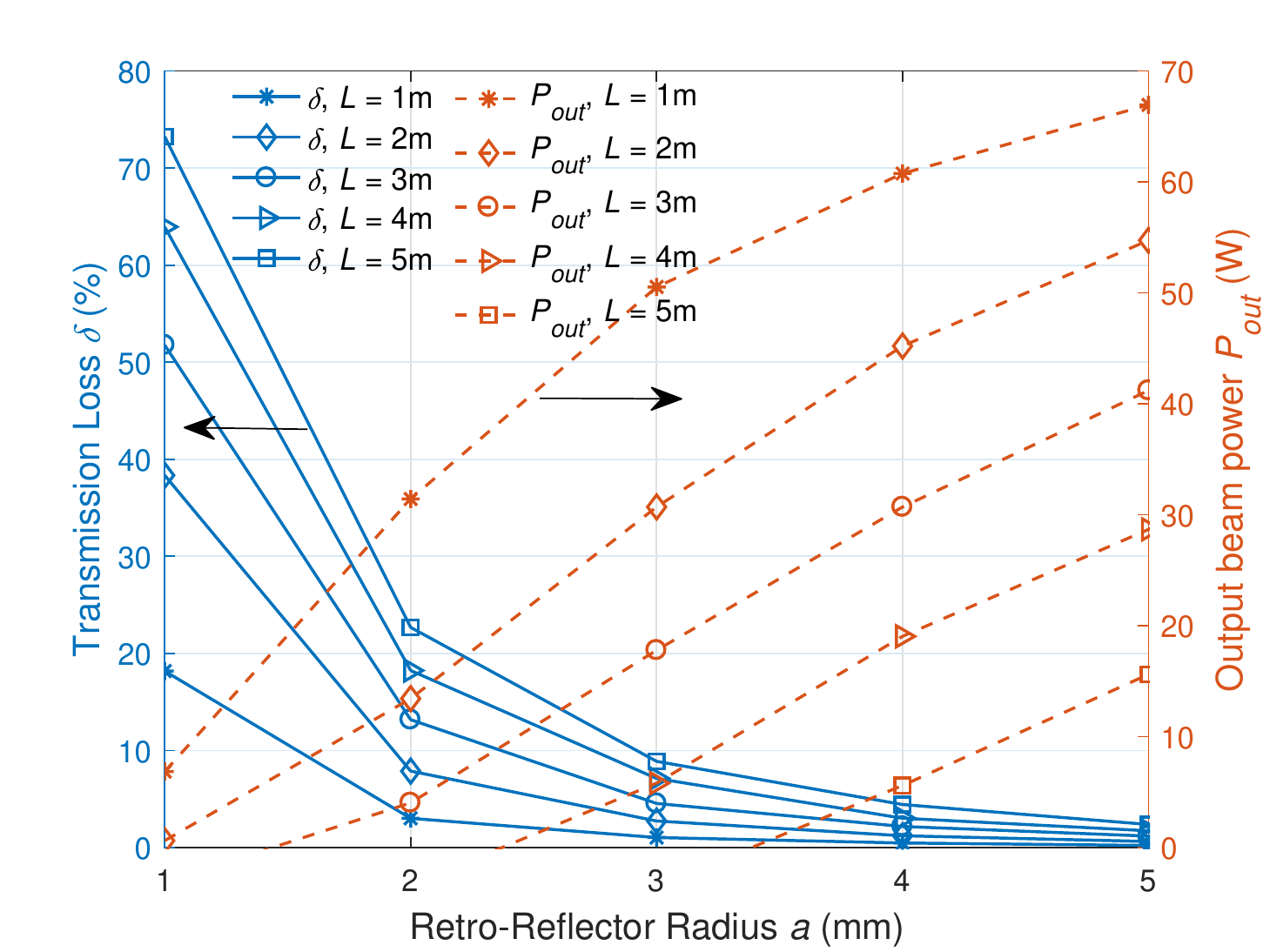}
    \caption{Transmission loss $\delta$ and output beam power $P_{out}$ versus retro-reflector radius $a$.}
    \label{Fig10}
\end{figure}

The beam transmission loss from transmitter to receiver in the RB-SWIPT system can be figured out based on intra-cavity power intensity calculated by \eqref{eq5} and transmission loss in \eqref{eq16}. The output beam power can be calculated based on \eqref{eq20}. First of all, the transmission loss $\delta$ and output beam power $P_{out}$ vary with cavity length $L$ as depicted in Fig. \ref{Fig9}. Since the beam power intensity decreases due to scattering and absorption inside the transmission medium, the transmission loss will increase with the increase of the transmission distance. As shown in Fig. \ref{Fig9}, because of the increased influence of the transmission environment, increasing cavity length $L$ leads that transmission loss $\delta$ amplifies under the radius of retro-reflector changing from $1mm$ to $5mm$. That is, the growth of the transmission distance will increase the transmission loss in the resonant cavity. If the radius of retro-reflector is $5mm$ and the cavity length is $1-5m$, the transmission loss is $0.21\%$, $0.62\%$, $1.17\%$, $1.72\%$, and $2.38\%$, respectively.

Conversely, due to the increased transmission loss, the output beam power $P_{out}$ will decrease with the growth of cavity length $L$. As depicted in Fig. \ref {Fig9}, $P_{out}$ decreases as L increases with different retro-reflector radii. If $P_{out}$ is less than $0W$, the pumping power at the transmitter does not reach the pumping threshold of the system, and there is no beam output. The output beam power is $66.94W$, $54.73W$, $41.20W$, $28.68W$, and $15.61W$ respectively if the radius of retro-reflector is $5mm$ and the cavity length is $1-5m$.

Furthermore, from Fig. \ref{Fig9}, once the cavity length $L$ is fixed, the transmission loss $\delta$ decreases and the output beam power $P_{out}$ increases when using input and output retro-reflectors with larger radius. Figure \ref{Fig10} depicts the change trends of transmission loss $\delta$ and output beam power $P_{out}$ as the variation of the retro-reflector radius $a$.

In the resonant cavity, the resonant beam is paraxial distribution, i.e. the beam energy is concentrated towards the optical axis. In Fig. \ref{Fig10}, since the power reception area increases, more resonant beam energy is stored in the cavity. Thus, the transmission loss $\delta$ decreases with retro-reflector radius $a$ getting larger. If the cavity length is $3m$, the transmission loss $\delta$ is about $51.76\%$, $13.17\%$, $4.54\%$, $2.17\%$, and $1.17\%$ respectively as the retro-reflector radius $a$ increases from $1mm$ to $5mm$. Similarly, the output beam power $P_{out}$ increases as $a$ increases. If the cavity length is $1m$, the output beam power is $6.85W$, $31.42W$, $50.53W$, $60.76W$, and $66.94W$ as $a$ is $1-5mm$.

\begin{figure}[!t]
    \centering
     \includegraphics[scale=0.58]{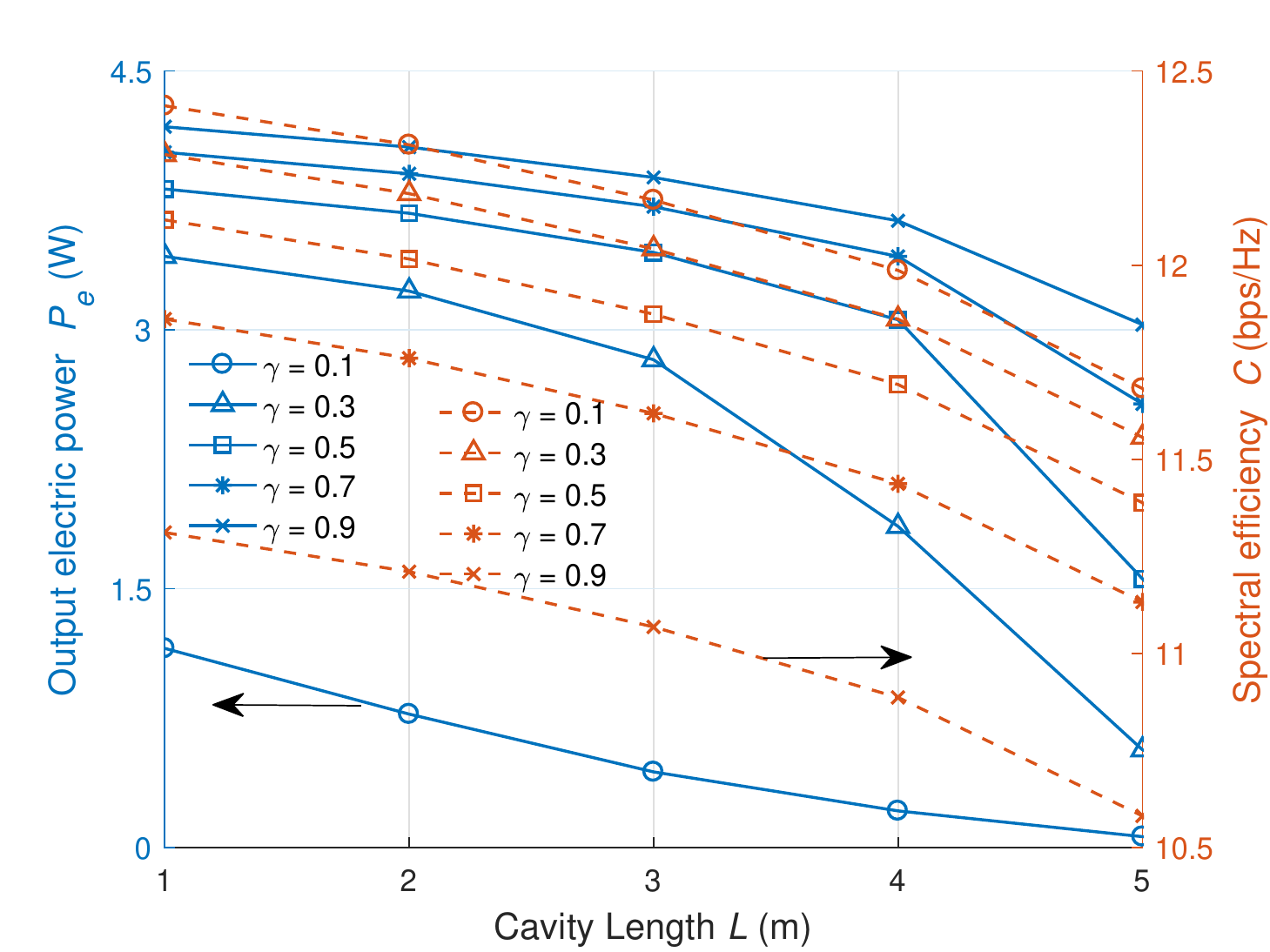}
    \caption{Output electric power $P_e$ and spectral efficiency $C$ versus resonant cavity length $L$.}
    \label{Fig13}
\end{figure}

\subsection{Energy Harvesting and Information Receiving}
According to \eqref{outelepower} and \eqref{SE}, the output electric power and the spectral efficiency can be calculated when the retro-reflector size and the transmission distance are known. Thus, we analyze the output electric power $P_e$ and the spectral efficiency $C$ here with the changes of cavity length $L$ and retro-reflector radius $a$ under different power-splitting ratio $\gamma$.

According to above analysis, if the radius of the retro-reflector is $5mm$, we can obtain the maximum output beam power. Here, if $a$ is $5mm$, we analyze the changes of output electric power $P_e$ and the spectral efficiency $C$ for power-splitting ratio $\gamma$ as $0.1$, $0.3$, $0.5$, $0.7$ and $0.9$ with the cavity length $L$ as shown in Fig. \ref{Fig13}. Based on \eqref{current} and \eqref{SP}, the output electric power $P_e$ and spectral efficiency $C$ are proportional to the output beam power $P_{out}$. As depicted in Fig. \ref{Fig13}, since the output beam power decreases as the cavity length $L$ extends, the output electric power $P_e$ and spectral efficiency $C$ decreases with the increase of $L$. For example, If $\gamma$ is $0.3$, $P_e$ is about $3.43W$, $3.22W$, $2.83W$, $1.86W$, and $0.57W$, and $C$ is $12.28bps/Hz$, $12.18bps/Hz$, $12.04bps/Hz$, $11.86bps/Hz$, and $11.57bps/Hz$ respectively when $L$ is $1-5m$.

\begin{figure}[!t]
    \centering
     \includegraphics[scale=0.58]{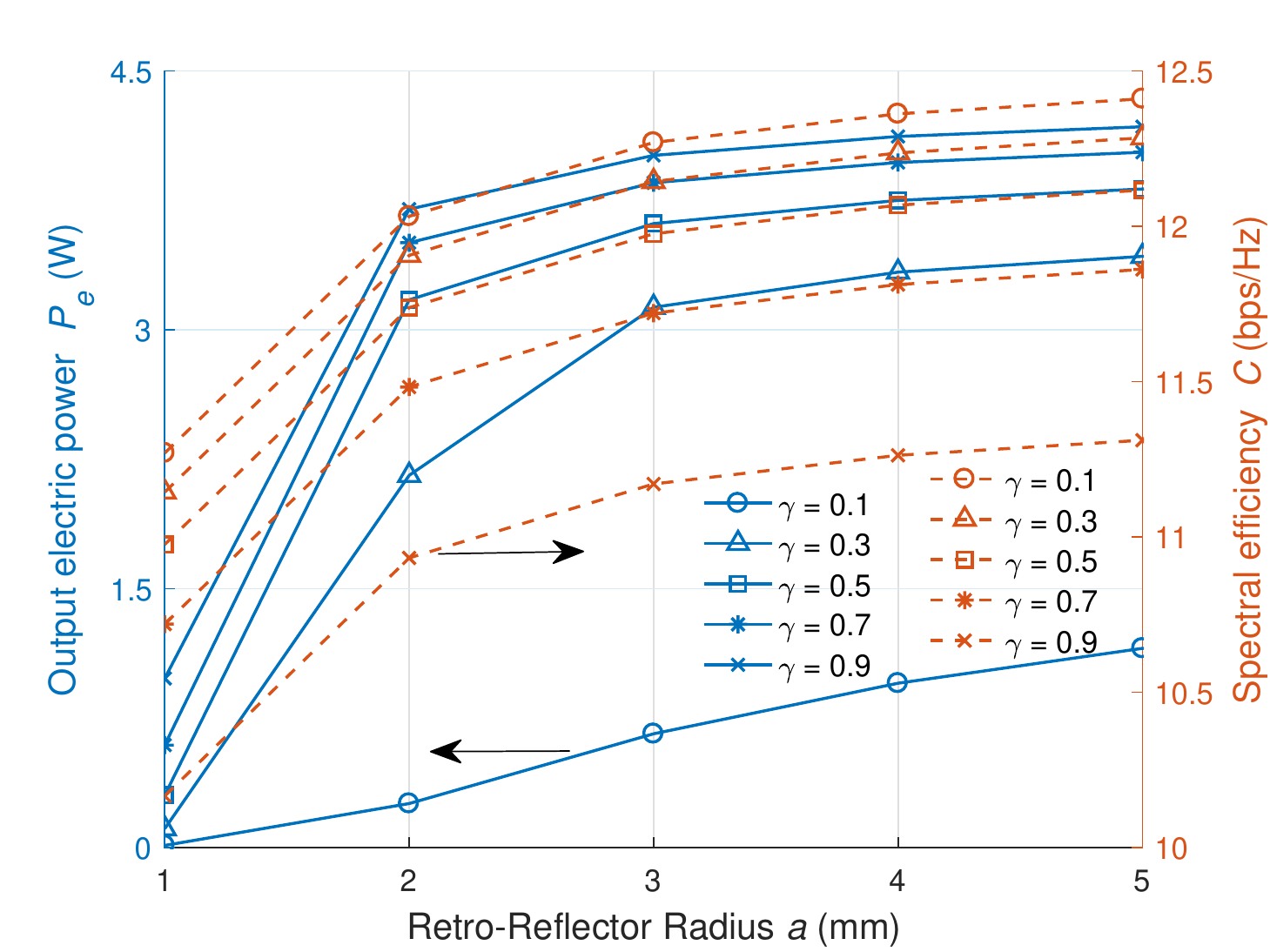}
    \caption{Output electric power $P_e$ and spectral efficiency $C$ versus retro-reflector radius $a$.}
    \label{Fig14}
\end{figure}

In addition, if the cavity length $L$ is $1m$, the maximum output beam power can be obtained in the RB-SWIPT system and the output beam power increases as the retro-reflector radius increases in Fig. \ref{Fig10}. The change trends of the output electric power $P_e$ and spectral efficiency $C$ with the radius of retro-reflector $a$ are demonstrated in Fig. \ref{Fig14}. It can be seen from Fig. \ref{Fig14} that $P_e$ and $C$ increase when $a$ gets larger independent of the power-splitting ratio $\gamma$. If $\gamma$ is $0.3$ and $a$ increases from $1mm$ to $5mm$, the output electric power $P_e$ is $0.10W$, $2.16W$, $3.13W$, $3.33W$, and $3.43W$, and spectral efficiency $C$ is $11.14bps/Hz$, $11.91bps/Hz$, $12.14bps/Hz$, $12.24bps/Hz$, and $12.28bps/Hz$, respectively.

\subsection{End-to-End Efficiency}
To analyze the factors affecting the efficiency of the RB-SWIPT system, we study the changing trend of end-to-end efficiency as the cavity length and retro-reflector radius vary based on the above analysis of transmission loss, output beam power, output electric power, and spectral efficiency.

Under the premise that the output beam power $P_{out}$ and the pumping optical power $P_{gb}$ are known, the end-to-end efficiency $\eta$ can be calculated using the ratio of $P_{out}$ to $P_{gb}$ (i.e. $P_{out}/P_{gb}$). According to \cite{crump2007100}, the electro-optical conversion efficiency is about $71.5\%$. Thus, $P_{gb}$ is constant in an RB-SWIPT system with fixed input electric power. In this paper, we study the RB-SWIPT performance with $200W$ input electric power. Therefore, $P_{gb}$ is about $143W$.

\begin{figure}[!t]
    \centering
     \includegraphics[scale=0.58]{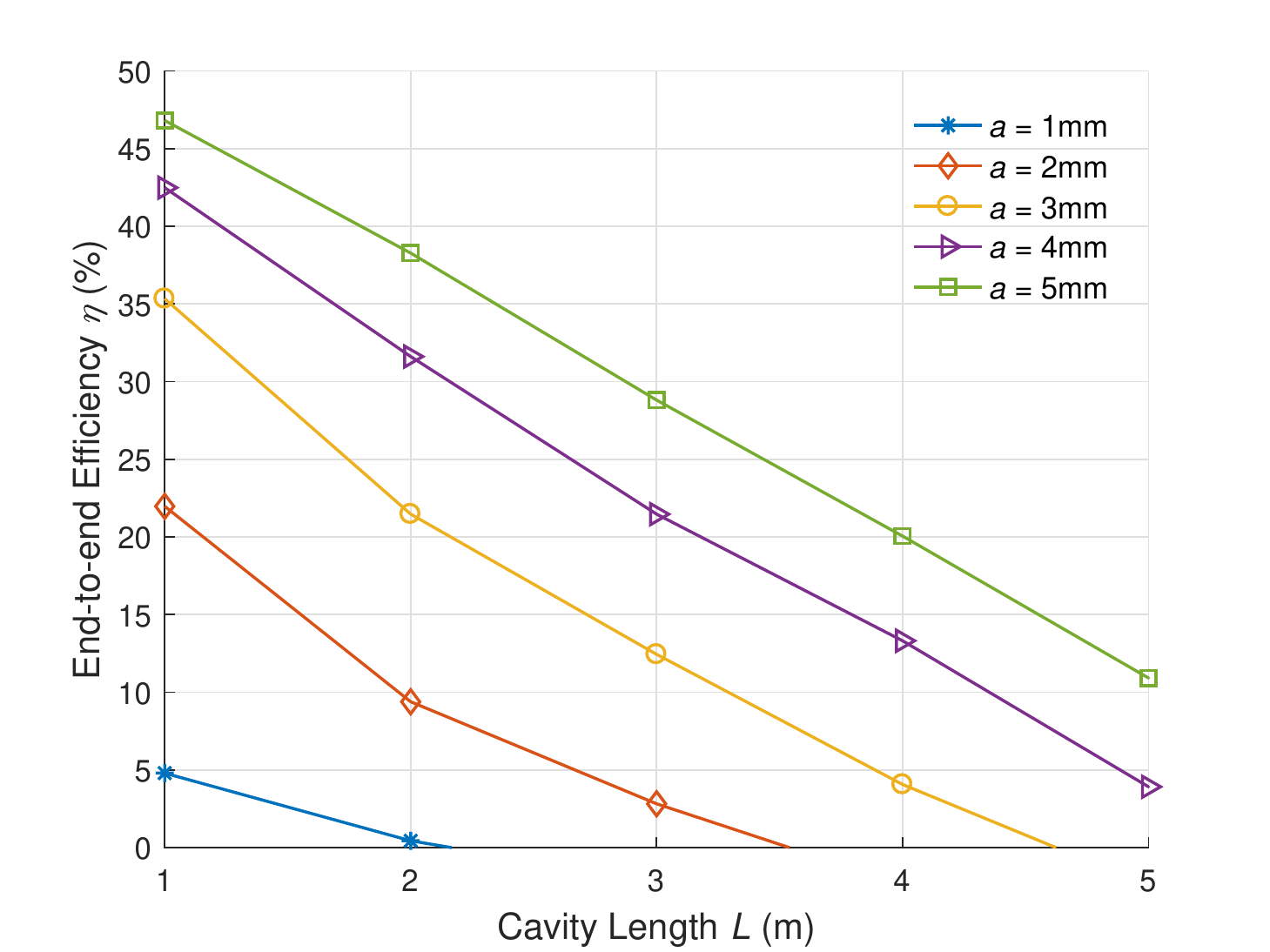}
    \caption{End-to-end efficiency $\eta$ versus cavity length $L$.}
    \label{Fig15}
\end{figure}

Since the output beam power $P_{out}$ in Figs. \ref{Fig9} and \ref{Fig10} decreases with the extension of the cavity length $L$, the end-to-end efficiency $\eta$ has the same trend as shown in Fig. \ref{Fig15}. No matter how much the radius is, the end-to-end efficiency decreases with the increase of cavity length. If the retro-reflector radius is $5mm$, $\eta$ is about $46.81\%$, $38.28\%$, $28.81\%$, $20.06\%$, and $10.92\%$ as the cavity length $L$ increases from $1m$ to $5m$.

\begin{figure}[!t]
    \centering
     \includegraphics[scale=0.58]{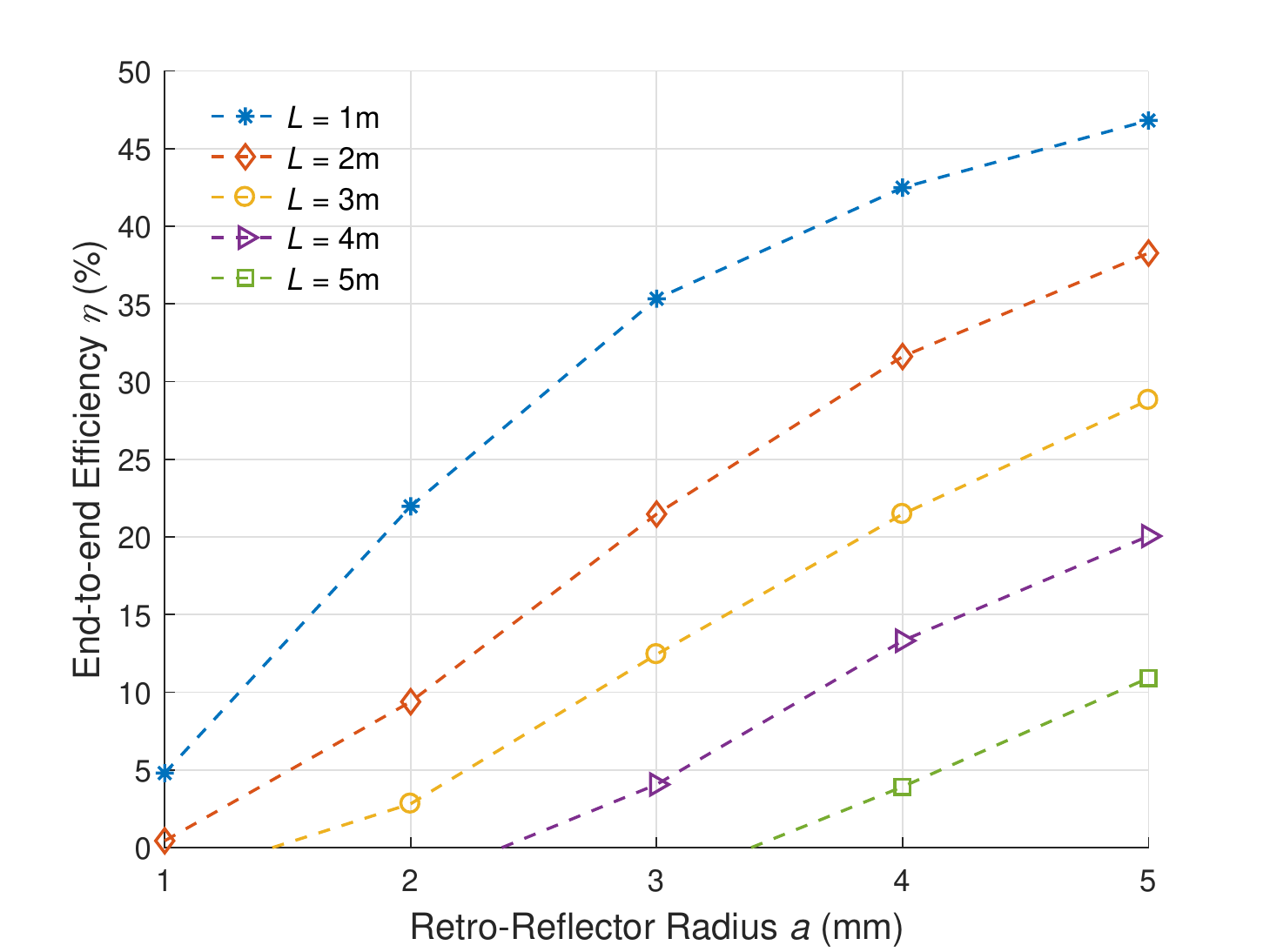}
    \caption{End-to-end efficiency $\eta$ versus retro-reflector radius $a$.}
    \label{Fig16}
\end{figure}

In addition, as shown in Fig. \ref{Fig15}, if the cavity length $L$ is fixed, the end-to-end efficiency $\eta$ increases as the radius of retro-reflector $a$ magnifies because the output beam power $P_{out}$ increases with the increase of retro-reflector radius $a$. The direct relationship between retro-reflector radius $a$ and the end-to-end efficiency $\eta$ can be obtained from Fig. \ref{Fig16}. $\eta$ has the same variation trend with $P_{out}$ as the retro-reflector radius $a$ changes. With the growth of the retro-reflector radius, the end-to-end efficiency rises regardless of the change of cavity length. That is, the larger radius $a$ brings the smaller loss $\delta$, leading the greater output beam power $P_{out}$, and finally causes the higher end-to-end efficiency $\eta$. When $L$ is $1m$, $\eta$ is $4.79\%$, $21.97\%$, $35.34\%$, $42.49\%$, and $46.81\%$ respectively as $a$ is $1-5mm$.

\subsection{Summary}
Based on the E$2$E SWIPT model of the RB-SWIPT system, the power intensity in the resonant cavity can be calculated and analyzed. Then, the transmission loss increases with the cavity length extending and decreases with the retro-reflector radius getting larger. Afterwards, the output beam power, output electric power, and spectral efficiency in the receiver increase with the decrease of cavity length and increase of retro-reflector radius. Afterwards, the end-to-end charging efficiency and the output beam power have the same change trend with the change in cavity length and retro-reflector radius. Moreover, if the cavity length is $1m$ and the retro-reflector radius is $5mm$, the maximum output electric power is about $4.20W$ ($\gamma = 0.9$), and the maximum spectral efficiency is $12.41bps/Hz$ approximately ($\gamma = 0.1$). Finally, the end-to-end efficiency is $46.81\%$ when the cavity length is $1m$ and the retro-reflector radius is $5mm$.

The requirements of energy supply and communication can be balanced by varying the power-splitting ratio. The operations to improve the E$2$E efficiency of the RB-SWIPT system include: i) designing a shorter resonant cavity (i.e. shorter transmission distance), and ii) using larger input/output retro-reflectors (i.e. a larger retro-reflector size).

\section{Conclusions and Future Work}\label{Section5}
The resonant beam simultaneous wireless information and power transfer (RB-SWIPT) system based on spatially separated laser-cavity is proposed in this paper. To reveal the energy-information transmission mechanism, we establish the end-to-end SWIPT model for calculating the intra-cavity power intensity, transmission loss, output electric power and spectral efficiency with different system parameters. Finally, the SWIPT performance is evaluated numerically with the $1-5m$ cavity length and $1-5mm$ retro-reflector radius. Numerical results illustrate that shorter transmission distance or larger retro-reflector sizes in both transmitter and receiver can improve the E$2$E efficiency for the RB-SWIPT system.

In the future research on the RB-SWIPT system, there are still many issues to be studied, for example:
\begin{enumerate}
  \item [\bf i1)] From electromagnetic field propagation for the RB-SWIPT system, the beam power intensity at any position in the resonant cavity can be obtained. Then, the safety of RB-SWIPT system can be analyzed based on the value of power intensity in the resonant cavity; and,
  \item [\bf i2)] The beam incident on a retro-reflector can be returned to the original path regardless of the angle of incidence. Afterwards, we can study the mobile energy transfer and communication using resonant beam.
\end{enumerate}

\bibliographystyle{IEEEtran}
\bibliographystyle{unsrt}
\bibliography{references}

\begin{thebibliography}{10}
\providecommand{\url}[1]{#1}
\csname url@samestyle\endcsname
\providecommand{\newblock}{\relax}
\providecommand{\bibinfo}[2]{#2}
\providecommand{\BIBentrySTDinterwordspacing}{\spaceskip=0pt\relax}
\providecommand{\BIBentryALTinterwordstretchfactor}{4}
\providecommand{\BIBentryALTinterwordspacing}{\spaceskip=\fontdimen2\font plus
\BIBentryALTinterwordstretchfactor\fontdimen3\font minus
  \fontdimen4\font\relax}
\providecommand{\BIBforeignlanguage}[2]{{%
\expandafter\ifx\csname l@#1\endcsname\relax
\typeout{** WARNING: IEEEtran.bst: No hyphenation pattern has been}%
\typeout{** loaded for the language `#1'. Using the pattern for}%
\typeout{** the default language instead.}%
\else
\language=\csname l@#1\endcsname
\fi
#2}}
\providecommand{\BIBdecl}{\relax}
\BIBdecl

\bibitem{chen2010improved}
Y.~Chen, ``Improved energy detector for random signals in gaussian noise,''
  \emph{IEEE Trans. Wireless Commun.}, vol.~9, no.~2, pp. 558--563, Feb. 2010.

\bibitem{krikidis2014simultaneous}
I.~Krikidis, S.~Timotheou, S.~Nikolaou, G.~Zheng, D.~W.~K. Ng, and R.~Schober,
  ``Simultaneous wireless information and power transfer in modern
  communication systems,'' \emph{IEEE Commun. Mag.}, vol.~52, no.~11, pp.
  104--110, Nov. 2014.

\bibitem{varshney2008transporting}
L.~R. Varshney, ``Transporting information and energy simultaneously,'' in
  \emph{2008 IEEE International Symposium on Information Theory}, Toronto,
  Ontario, Canada, July 6-11 2008, pp. 1612--1616.

\bibitem{agiwal2016next}
M.~Agiwal, A.~Roy, and N.~Saxena, ``Next generation 5{G} wireless networks: A
  comprehensive survey,'' \emph{IEEE Communications Surveys \& Tutorials},
  vol.~18, no.~3, pp. 1617--1655, Feb.

\bibitem{perera2018simultaneous}
T.~D.~P. Perera, D.~N.~K. Jayakody, S.~K. Sharma, S.~Chatzinotas, and J.~Li,
  ``Simultaneous wireless information and power transfer ({SWIPT}): Recent
  advances and future challenges,'' \emph{IEEE Communications Surveys and
  Tutorials}, vol.~20, no.~1, pp. 264--302, 2018.

\bibitem{liu2020big}
X.~Liu, Q.~Sun, W.~Lu, C.~Wu, and H.~Ding, ``Big-data-based intelligent
  spectrum sensing for heterogeneous spectrum communications in 5{G},''
  \emph{IEEE Wireless Communications}, vol.~27, no.~5, pp. 67--73, Oct. 2020.

\bibitem{masotti2021rf}
D.~Masotti, M.~Shanawani, G.~Murtaza, G.~Paolini, and A.~Costanzo, ``{RF}
  systems design for simultaneous wireless information and power transfer
  ({SWIPT}) in automation and transportation,'' \emph{IEEE Journal of
  Microwaves}, vol.~1, no.~1, pp. 164--175, Jan. 2021.

\bibitem{liu2016swipt}
W.~Liu, X.~Zhou, S.~Durrani, and P.~Popovski, ``{SWIPT} with practical
  modulation and {RF} energy harvesting sensitivity,'' in \emph{2016 IEEE
  International Conference on Communications (ICC)}, Kuala Lumpur, Malaysia,
  May 23-27 2016, pp. 1--7.

\bibitem{ma2019simultaneous}
S.~Ma, F.~Zhang, H.~Li, F.~Zhou, Y.~Wang, and S.~Li, ``Simultaneous lightwave
  information and power transfer in visible light communication systems,''
  \emph{IEEE Trans. Wireless Commun.}, vol.~18, no.~12, pp. 5818--5830, Dec.
  2019.

\bibitem{fakidis20180}
J.~Fakidis, S.~Videv, H.~Helmers, and H.~Haas, ``0.5-{G}b/s {OFDM}-based laser
  data and power transfer using a {G}a{A}s photovoltaic cell,'' \emph{IEEE
  Photon. Technol. Lett.}, vol.~30, no.~9, pp. 841--844, May 2018.

\bibitem{liu2016dlc}
Q.~Liu, J.~Wu, P.~Xia, S.~Zhao, W.~Chen, Y.~Yang, and L.~Hanzo, ``Charging
  unplugged: {W}ill distributed laser charging for mobile wireless power
  transfer work?'' \emph{IEEE Veh. Technol. Mag.}, vol.~11, no.~4, pp. 36--45,
  Dec. 2016.

\bibitem{fang2018}
W.~Fang, Q.~Zhang, Q.~Liu, J.~Wu, and P.~Xia, ``Fair scheduling in resonant
  beam charging for {I}o{T} devices,'' \emph{IEEE Internet Things J.}, vol.~6,
  no.~1, pp. 641--653, Feb. 2019.

\bibitem{zhang2018distributed2}
Q.~Zhang, W.~Fang, Q.~Liu, J.~Wu, P.~Xia, and L.~Yang, ``Distributed laser
  charging: {A} wireless power transfer approach,'' \emph{IEEE Internet Things
  J.}, vol.~5, no.~5, pp. 3853--3864, Oct. 2018.

\bibitem{wang2018channel}
W.~Wang, Q.~Zhang, H.~Li, M.~Liu, X.~Liao, and Q.~Liu, ``Wireless energy
  transmission channel modeling in resonant beam charging for {I}o{T}
  devices,'' \emph{IEEE Internet Things J.}, vol.~6, no.~2, pp. 3976--3986,
  Apr. 2019.

\bibitem{david20186g}
K.~David and H.~Berndt, ``6{G} vision and requirements: Is there any need for
  beyond 5{G}?'' \emph{IEEE Veh. Technol. Mag.}, vol.~13, no.~3, pp. 72--80,
  Sept. 2018.

\bibitem{xiong2019resonant}
M.~Xiong, Q.~Liu, G.~Wang, G.~B. Giannakis, and C.~Huang, ``Resonant beam
  communications: Principles and designs,'' \emph{IEEE Commun. Mag.}, vol.~57,
  no.~10, pp. 34--39, Oct. 2019.

\bibitem{chen2019resonant}
W.~Chen, S.~Zhao, Q.~Shi, and R.~Zhang, ``Resonant beam charging-powered
  {UAV}-assisted sensing data collection,'' \emph{IEEE Trans. Veh. Technol.},
  vol.~69, no.~1, pp. 1086--1090, Jan. 2020.

\bibitem{hodgson2005laser}
N.~Hodgson and H.~Weber, \emph{Laser Resonators and Beam Propagation:
  Fundamentals, Advanced Concepts, Applications}.\hskip 1em plus 0.5em minus
  0.4em\relax Springer, 2005.

\bibitem{lim2019wireless}
J.~Lim, T.~S. Khwaja, and J.~Ha, ``Wireless optical power transfer system by
  spatial wavelength division and distributed laser cavity resonance,''
  \emph{Optics express}, vol.~27, no.~12, pp. A924--A935, June 2019.

\bibitem{rabinovich2004performance}
W.~S. Rabinovich, P.~G. Goetz, R.~Mahon, L.~Swingen, J.~Murphy, G.~C.
  Gilbreath, S.~C. Binari, and E.~Waluschka, ``Performance of cat's eye
  modulating retroreflectors for free-space optical communications,'' in
  \emph{Free-Space Laser Communications IV. International Society for Optics
  and Photonics}, vol. 5550, Oct. 2004, pp. 104--114.

\bibitem{snyder1975paraxial}
J.~Snyder, ``Paraxial ray analysis of a cat's-eye retroreflector,''
  \emph{Applied optics}, vol.~14, no.~8, pp. 1825--1828, Aug. 1975.

\bibitem{poirson1997analytical}
J.~Poirson, F.~Bretenaker, M.~Vallet, and A.~Le~Floch, ``Analytical and
  experimental study of ringing effects in a {F}abry--{P}erot cavity.
  application to the measurement of high finesses,'' \emph{Journal of the
  Optical Society of America B}, vol.~14, no.~11, pp. 2811--2817, Nov. 1997.

\bibitem{eftekharnejad2012impact}
S.~Eftekharnejad, V.~Vittal, G.~T. Heydt, B.~Keel, and J.~Loehr, ``Impact of
  increased penetration of photovoltaic generation on power systems,''
  \emph{IEEE Trans. Power Syst.}, vol.~28, no.~2, pp. 893--901, May 2013.

\bibitem{fox1963modes}
A.~Fox and T.~Li, ``Modes in a maser interferometer with curved and tilted
  mirrors,'' \emph{Proc. IEEE}, vol.~51, no.~1, pp. 80--89, Jan. 1963.

\bibitem{lissak1990transverse}
B.~Lissak and S.~Ruschin, ``Transverse pattern modifications in a stable
  apertured laser resonator,'' \emph{Applied optics}, vol.~29, no.~6, pp.
  767--771, Feb. 1990.

\bibitem{fox1961resonant}
A.~G. Fox and T.~Li, ``Resonant modes in a maser interferometer,'' \emph{Bell
  System Technical Journal}, vol.~40, no.~2, pp. 453--488, Mar. 1961.

\bibitem{gordon1964equivalence}
J.~Gordon and H.~Kogelnik, ``Equivalence relations among spherical mirror
  optical resonators,'' \emph{Bell System Technical Journal}, vol.~43, no.~6,
  pp. 2873--2886, Nov. 1964.

\bibitem{li1965diffraction}
T.~Li, ``Diffraction loss and selection of modes in maser resonators with
  circular mirrors,'' \emph{Bell System Technical Journal}, vol.~44, no.~5, pp.
  917--932, May 1965.

\bibitem{sera2007pv}
D.~Sera, R.~Teodorescu, and P.~Rodriguez, ``P{V} panel model based on datasheet
  values,'' in \emph{2007 IEEE International Symposium on Industrial
  Electronics}, Vigo, Spain, 4-7 June 2007, pp. 2392--2396.

\bibitem{cvijetic2008performance}
N.~Cvijetic, S.~G. Wilson, and M.~Brandtpearce, ``Performance bounds for
  free-space optical {MIMO} systems with {APD} receivers in atmospheric
  turbulence,'' \emph{IEEE J. Sel. Areas Commun.}, vol.~26, no.~3, pp. 3--12,
  Apr. 2008.

\bibitem{lapidoth2009on}
A.~Lapidoth, S.~M. Moser, and M.~Wigger, ``On the capacity of free-space
  optical intensity channels,'' \emph{IEEE Trans. Inf. Theory}, vol.~55,
  no.~10, pp. 4449--4461, Nov. 2009.

\bibitem{perales2016characterization}
M.~Perales, M.-h. Yang, C.-l. Wu, C.-w. Hsu, W.-s. Chao, K.-h. Chen, and
  T.~Zahuranec, ``Characterization of high performance silicon-based {VMJ} {PV}
  cells for laser power transmission applications,'' in \emph{High-Power Diode
  Laser Technology and Applications XIV}, vol. 9733, Mar. 2016, p. 97330U.

\bibitem{crump2007100}
P.~Crump, W.~Dong, M.~Grimshaw, J.~Wang, S.~Patterson, D.~Wise, M.~Defranza,
  S.~Elim, S.~Zhang, M.~Bougher \emph{et~al.}, ``100-{W}+ diode laser bars show
  $>$ 71\% power conversion from 790- nm to 1000-nm and have clear route to $>$
  85\%,'' \emph{Proceedings of SPIE}, vol. 6456, Feb. 2007.

\bibitem{demir2017handover}
M.~S. Demir, F.~Miramirkhani, and M.~Uysal, ``Handover in {VLC} networks with
  coordinated multipoint transmission,'' in \emph{2017 IEEE International Black
  Sea Conference on Communications and Networking (BlackSeaCom)}, Istanbul,
  Turkey, Jun. 5-8 2017, pp. 1--5.

\bibitem{quintana2017high}
C.~Quintana, Q.~Wang, D.~Jakonis, X.~Piao, G.~Erry, D.~Platt, Y.~Thueux,
  A.~Gomez, G.~Faulkner, H.~Chun, M.~Salter, and O.~D, ``High speed
  electro-absorption modulator for long range retroreflective free space
  optics,'' \emph{IEEE Photon. Technol. Lett.}, vol.~29, no.~9, pp. 707--710,
  Mar. 2017.

\bibitem{moreira1997optical}
A.~J. Moreira, R.~T. Valadas, and A.~de~Oliveira~Duarte, ``Optical interference
  produced by artificial light,'' \emph{Wirel. Netw.}, vol.~3, no.~2, pp.
  131--140, May 1997.

\bibitem{xu2011impact}
F.~Xu, M.-A. Khalighi, and S.~Bourennane, ``Impact of different noise sources
  on the performance of {PIN}-and {APD}-based {FSO} receivers,'' in
  \emph{Proceedings of the 11th International Conference on
  Telecommunications}, Graz, Austria, Jun. 15-17 2011, pp. 211--218.

\end{thebibliography}

\end{document}